\documentclass[11pt,a4paper,preprintnumbers,nofootinbib]{revtex4}

\pdfoutput=1

\usepackage[T1]{fontenc}
\usepackage[utf8]{inputenc}
\usepackage{amsmath}
\usepackage{amssymb}
\usepackage{hyperref}
\usepackage[final]{graphicx}

\begin{document}

\title{Flavor \& new physics opportunities with  rare charm decays into leptons}
\author{Stefan de Boer}
\email{stefan.deboer@tu-dortmund.de}
\author{Gudrun Hiller}
\email{ghiller@physik.uni-dortmund.de}
\affiliation{Fakultät für Physik, TU Dortmund, Otto-Hahn-Str.4, D-44221 Dortmund, Germany}
\begin{abstract}
Updated standard model predictions for $D_{(s)} \to P l^+ l^-$, $ l=e, \mu$, $P=\pi,K$ and inclusive decays are presented.
Model-independent constraints on $|\Delta C|=|\Delta U|=1$ Wilson coefficients are worked out.
New physics (NP) opportunities do arise in semileptonic branching ratios for very large couplings only, however, are not excluded outside the resonance regions yet.
The NP potential of resonance-assisted CP-asymmetries and angular observables  is worked out.
Predictions are given  for leptoquark models,  and include lepton flavor violating and dineutrino decays.
Whether NP can be seen depends on flavor patterns, and vice versa.
\end{abstract}

\preprint{DO-TH 15/10, QFET-2015-25}

\maketitle

\section{Introduction}

The study of flavor-changing neutral current (FCNC) transitions is a key tool to explore the generational structure of standard model (SM) fermions and to look for physics beyond the standard model (BSM).
While analyses involving $b$-quarks are matured to precision level \cite{Blake:2015tda}, the investigation of charm  FCNCs is much less advanced,
as corresponding rates are highly GIM-suppressed and experimentally challenging  and/or decay modes subjected to resonance contributions, shielding the electroweak physics.

Semileptonic charm hadron decays  provide  an opportunity to probe  for new physics in $|\Delta C|=|\Delta U|=1$ FCNCs \cite{Hewett:2004tv}.
Such processes, induced by $c \to u l^+  l^-$, $l=e,\mu$, allow to kinematically reduce the resonance background via $ c\to M u \to  l^+ l^- u$, where $M$  denotes a meson with mass $m_M$ decaying to dileptons such as $M=\eta^{(\prime)}, \rho, \omega, \phi$, by kinematic cuts in the dilepton invariant mass squared $q^2$, notably $q^2 \gtrsim m_\phi^2$.
The available phase space is, however, limited, at most $\Delta q^2 \sim 2  \, \mbox{GeV}^2$ for the most favorable decays $D^+ \to \pi^+ l^+ l^-$, and the resonance tails remain overwhelming  in the decay rates until the endpoint. 
 To access short-distance physics becomes still possible
in  two situations: {\it i)}  The BSM-induced rates are much larger than the SM background.
{\it ii)}  Using SM null tests, that is, specifically chosen observables. The latter are generically related to SM (approximate) symmetries, such as CP  in $c \to u$ transitions, and include various ratios and asymmetries.

 In this work we pursue the analysis of rare charm observables using CP-asymmetries and 
 those related to leptons,  lepton flavor violating (LFV) decays
 $c \to u e^\pm \mu^\mp$. The latter have essentially no SM contribution due to the smallness of neutrino masses. Importantly, there are no
 photon-induced dilepton effects, the usual source of resonance pollution. Therefore, for LFV charm decays no cuts in $q^2$ are required from the theory perspective.
Similarly $c \to u \nu \bar \nu$ processes  have essentially zero  SM  background and factorize in the full region of $q^2$.
In addition, the study of rare charm decays has great prospects at the LHCb and Belle II experiments, as well as  BES III \cite{Asner:2008nq}, and  possible other  future high luminosity flavor facilities \cite{Aushev:2010bq,O'Leary:2010af}.

Leptoquarks are particularly interesting for flavor physics because they link quark flavor to lepton flavor. A rich phenomenology and correlations between 
different kinds of flavor transitions, $K$-, $D$- and $B$-decays as well as LFV, allow to
probe the SM and flavor models simultaneously.
Naturally, CP-violation, lepton-nonuniversality (LNU)  and LFV arise.
We work out  correlations in a number of flavor benchmarks for scalar and vector leptoquarks that induce $c \to u l^+l^-$.
Some of these are currently discussed in the context of $B$-physics anomalies  hinting at LNU \cite{Dorsner:2013tla,Hiller:2014yaa,Gripaios:2014tna,Calibbi:2015kma,Freytsis:2015qca}.

The plan of the paper is as follows:
In Sec.~\ref{sec:SM} we work out SM predictions for $c \to u  l^+ l^-$ processes, including recent results for higher order perturbative contributions  \cite{deBoerSeidel}.
We identify  BSM windows in rare exclusive $c \to u l^+ l^-$ modes.
In Sec.~\ref{sec:BSM} 
constraints and predictions are worked out model-independently and 
within leptoquark scenarios,  amended by flavor patterns.
In Sec.~\ref{sec:summary} we summarize.
Auxiliary information is compiled in several appendices.

\section{Standard Model predictions \label{sec:SM}}

We work out SM predictions for  exclusive semileptonic charm decays. In Sec.~\ref{sec:wilson_coefficients} we obtain (Next-to-)Next-to-Leading-Order (N)NLO results for the
(effective) $|\Delta C|=|\Delta U|=1$ coefficients.
In Sec.~\ref{sec:pheno} we work out branching ratios, including resonance effects and compare to data.

\subsection{Wilson Coefficients}\label{sec:wilson_coefficients}

We write the $c\to u l^+ l^-$ effective weak Lagrangian \cite{Greub:1996wn,Fajfer:2002gp,deBoerSeidel} with  two-step matching at  the $W$-mass scale and the $b$-quark mass scale, respectively,  as
\begin{align}
 &\mathcal L_\text{eff}^\text{weak}\big|_{m_W\ge\mu>m_b}=\frac{4G_F}{\sqrt 2}\sum_{q=d,s,b}V_{cq}^*V_{uq}\left(\tilde C_1(\mu)P_1^{(q)}(\mu)+\tilde C_2(\mu)P_2^{(q)}(\mu)\right)\,,\label{eq:L_eff_weak_mW}\\
 &\mathcal L_\text{eff}^\text{weak}\big|_{m_b>\mu\ge m_c}=\frac{4G_F}{\sqrt 2}\sum_{q=d,s}V_{cq}^*V_{uq}\left(\tilde C_1(\mu)P_1^{(q)}(\mu)+\tilde C_2(\mu)P_2^{(q)}(\mu) + \sum_{i=3}^{10}\tilde C_i(\mu)P_i(\mu)\right)\,.\label{eq:L_eff_weak_mb}
\end{align}
Here, $G_F$ is the Fermi constant and $V_{ij}$ denote the Cabibbo–Kobayashi–Maskawa (CKM)  matrix elements.
Within the OPE (\ref{eq:L_eff_weak_mW}), (\ref{eq:L_eff_weak_mb}) heavy fields are integrated in the Wilson coefficients $\tilde C_i$ and the operators $P_i$ are composed out of  light fields.
The SM operators up to dimension six read \cite{Chetyrkin:1996vx,Bobeth:1999mk,Gambino:2003zm}
\begin{align}
 &P_1^{(q)}=(\bar u_L\gamma_{\mu_1}T^aq_L)(\overline q_L\gamma^{\mu_1}T^ac_L)\,,\\
 &P_2^{(q)}=(\bar u_L\gamma_{\mu_1}q_L)(\overline q_L\gamma^{\mu_1}c_L)\,,\\
 &P_3=(\bar u_L\gamma_{\mu_1}c_L)\sum_{\{q:m_q<\mu\}}(\overline q\gamma^{\mu_1}q)\,,\\
 &P_4=(\bar u_L\gamma_{\mu_1}T^ac_L)\sum_{\{q:m_q<\mu\}}(\overline q\gamma^{\mu_1}T^aq)\,,\\
 &P_5=(\bar u_L\gamma_{\mu_1}\gamma_{\mu_2}\gamma_{\mu_3}c_L)\sum_{\{q:m_q<\mu\}}(\overline q\gamma^{\mu_1}\gamma^{\mu_2}\gamma^{\mu_3}q)\,,\\
 &P_6=(\bar u_L\gamma_{\mu_1}\gamma_{\mu_2}\gamma_{\mu_3}T^ac_L)\sum_{\{q:m_q<\mu\}}(\overline q\gamma^{\mu_1}\gamma^{\mu_2}\gamma^{\mu_3}T^aq)\,,\\
 &P_7=\frac e{g_s^2}m_c\left(\bar u_L\sigma^{\mu_1\mu_2}c_R\right)F_{\mu_1\mu_2}\,,\\
 &P_8=\frac 1{g_s}m_c\left(\bar u_L\sigma^{\mu_1\mu_2}T^ac_R\right)G^a_{\mu_1\mu_2}\,,\\
 &P_9=\frac{e^2}{g_s^2}(\bar u_L\gamma_{\mu_1}c_L)\left(\overline l\gamma^{\mu_1}l\right)\,,\\
 &P_{10}=\frac{e^2}{g_s^2}(\bar u_L\gamma_{\mu_1}c_L)\left(\overline l\gamma^{\mu_1}\gamma_5l\right)\,,
\end{align}
where $q_{L,R}=(1\mp\gamma_5)/2\,q$ denotes chiral quark fields, $T^a$ are the $SU(3)_C$ generators, $e$ is the electromagnetic coupling, $g_s$ is the strong coupling, $\sigma^{\mu_1\mu_2}=i[\gamma^{\mu_1},\gamma^{\mu_2}]/2$, $F_{\mu_1\mu_2}$ is the electromagnetic field strength tensor, $G_{\mu_1\mu_2}^a$ is the chromomagnetic field strength tensor and the covariant derivative is $D_\mu=\partial_\mu+ig_s\mathcal A_\mu^aT^a+ieQA_\mu$.

In this section we  give results for the (N)NLO QCD SM Wilson coefficients
\begin{align} \label{eq:WCexp}
 \tilde C_i(\mu)=\tilde C_i^{(0)}(\mu)+\frac{\alpha_s(\mu)}{4\pi}\tilde C_i^{(1)}(\mu)+\left(\frac{\alpha_s(\mu)}{4\pi}\right)^2\tilde C_i^{(2)}(\mu)+\mathcal O\left(\alpha_s^3(\mu)\right)\,.
\end{align}
$\tilde C_{1,2}(\mu_W)$  can be inferred from \cite{Bobeth:1999mk,Gorbahn:2004my}
and $\tilde C_{3-10}(\mu_W)=0$ due to CKM unitarity for vanishing light quark masses.
If one were to keep finite light quark masses in
the Wilson coefficients at $\mu_W$  as in \cite{Burdman:2001tf,Paul:2011ar,Wang:2014uiz} spurious large logarithms are induced, {\it e.g.},  \cite{Inami:1980fz}
\begin{align}\label{eq:large_logarithm}
 \sum_{q=d,s,b}V_{cq}^*V_{uq}\tilde C_9^{(q)}(\mu_W)\simeq V_{cs}^*V_{us}\frac{(-2)}9\ln\frac{m_s^2}{m_d^2}\simeq-0.29\,,
\end{align}
a procedure 
that is not consistent with the factorization of scales in the effective Lagrangian  \cite{Fajfer:2002gp,deBoerSeidel}.
Logarithms are resummed to all orders in perturbation theory via the renormalization group (RG) equation \cite{Gorbahn:2004my,Czakon:2006ss,Gorbahn:2005sa}.
After RG-evolution of   $\tilde C_{1,2}$ from $\mu_W$ to $\mu_b$, we integrate out the $b$-quark at $\mu_b$, which induces non-zero contributions to $P_{3-10}$, and then RG-evolve 
$\tilde C_{1-10}$ from $\mu_b$ to $\mu_c$.
The resummation to NNLO  is worked out in \cite{deBoerSeidel}, to which we refer for details on the RG equation, anomalous dimensions and matching.
The results of this NNLO evolution are included in the numerical analysis in this work. 
Using the parameters compiled in App.~\ref{app:parameters} we find the SM Wilson coefficients at the charm quark mass given in Table~\ref{tab:WC}.

\begin{table}[!htb]
 \centering
 \begin{tabular}{c|c|c|c|c|c|c|c|c|c|c}
    \mbox{} &  $j=1$ &  $j=2$ &  $j=3$ &  $j=4$ & $j=5$ &  $j=6$ &  $j=7$ &  $j=8$ &  $j=9$ &  $j=10 $ \\
  \noalign{\hrule height 1pt}
  $\tilde C_j^{(0)}$                       &  -1.0275             &   1.0926           &  -0.0036             &  -0.0604             &   0.0004             &   0.0007             &   0                  &   0                  &  -0.0030             &  0  \\
  $(\alpha_s/(4\pi)) \, \tilde C_j^{(1)}$                        &   0.3214             &  -0.0549             &   -0.0025             &   -0.0312             &  0.0000             &  -0.0002             &   0.0035             &  -0.0020             &   -0.0064             &  0  \\
  $(\alpha_s/(4\pi))^2 \, \tilde C_j^{(2)}$                        &   0.0766             &  -0.0037             &  -0.0019             &   -0.0008             &   0.0001             &  0.0003             &   0.0002             &  -0.0003             &  -0.0037             &  0  \\
  \hline
  $\tilde C_j$  &  -0.6295  &  1.0340  &  -0.0080  &  -0.0924  &  0.0005  &  0.0008  &  0.0037  &  -0.0023  &  -0.0131  &  0  \\
 \end{tabular}
 \caption{The $i$th order contributions $(\alpha_s/(4\pi))^i\,\tilde C_j^{(i)}$, $i=0,1,2$  to the SM Wilson coefficients, see Eq.~(\ref{eq:WCexp}), at $\mu_c=m_c$. The last row gives their sum, 
 $\tilde C_j(m_c)$. }
  \label{tab:WC}
\end{table}

We write the matrix elements of the operators $P_{1-6,8}$ in terms of effective Wilson coefficients $C_{7,9}^\text{eff}(\mu_c)$ and $C_{10}^\text{eff}(\mu_c)=0$.
We find to one-loop order
\begin{align}\label{eq:C9_eff}
 C_9^{\text{eff}(q)}(q^2)&=\tilde C_9+\frac{\alpha_s}{4\pi}\bigg[ \frac8{27}\tilde C_1+\frac29\tilde C_2-\frac89\tilde C_3-\frac{32}{27}\tilde C_4-\frac{128}9\tilde C_5-\frac{512}{27}\tilde C_6\nonumber\\
 &+L(m_c^2,q^2)\left(\frac{28}{9}\tilde C_3+\frac{16}{27}\tilde C_4+\frac{304}9\tilde C_5+\frac{256}{27}\tilde C_6\right)\nonumber\\
 &+L(m_s^2,q^2)\left(-\frac43\tilde C_3-\frac{40}3\tilde C_5\right)\nonumber\\
 &+L(0,q^2)\left(\frac{16}9\tilde C_3+\frac{16}{27}\tilde C_4+\frac{184}9\tilde C_5+\frac{256}{27}\tilde C_6\right)\nonumber\\
 &+\left(\delta_{qs}L(m_s^2,q^2)+\delta_{qd}L(0,q^2)\right)\left(-\frac8{27}\tilde C_1-\frac29\tilde C_2\right)\bigg] \nonumber\\
 &+\left(\frac{\alpha_s}{4\pi}\right)^2F_8^{(9)}(q^2/m_c^2)C_8^\text{eff}\,,
\end{align}
in agreement with the corresponding calculation in $b$-quark decays \cite{Buras:1994dj} and
\begin{align}
 C_7^{\text{eff}(q)}(q^2)=\tilde C_7+\frac{\alpha_s}{4\pi}\sum_{i=1}^6y_i^{(7)}\tilde C_i+\left(\frac{\alpha_s}{4\pi}\right)^2 \! \! \left[\left(-\frac16\tilde C_1^{(0)}+\tilde C_2^{(0)}\right)f(m_q^2/m_c^2)+F_8^{(7)}(q^2/m_c^2)C_8^\text{eff}\right]
\end{align}
with $C_8^\text{eff}=\tilde C_8^{(1)}+\sum_{i=1}^6y_i^{(8)}\tilde C_i^{(0)}$ and $\tilde C_{1-6}$ consistently expanded to order $\alpha_s$.
The functions $f,$ $L$ and $F_8^{(7,9)}$ and the coefficients $y_i^{(7,8)}$ are given in App.~\ref{app:effective_wilson_coefficients}.
The coefficients $C_{7,9}^\text{eff}\sim V_{cb}^*V_{ub}$ induced by the two-loop matrix elements of $P_{3-6}$ and $C_9^\text{eff}$ induced by the two-loop matrix elements of $P_{1,2}$ are not known presently and neglected in the following analysis.
Hence, the NNLO result is not known; it is labeled as (N)NLO.

For the phenomenological analyses in Secs.~\ref{sec:pheno} and \ref{sec:BSM} it is customary to redefine the dilepton and electromagnetic dipole operators
and use 
\begin{align}
 &Q_7=\frac{m_c}e(\bar u\sigma^{\mu\nu}P_Rc)F_{\mu\nu}\,,\label{eq:Q7}\\
 &Q_9=(\bar u\gamma_{\mu}P_Lc)\left(\overline l\gamma^{\mu}l\right)\,, \label{eq:Q9}\\
 &Q_{10}=(\bar u\gamma_{\mu}P_Lc)\left(\overline l\gamma^{\mu}\gamma_5l\right)\label{eq:Q10} \, .
\end{align}
Their effective coefficients $C_{7,9,10}= C_{7,9,10}(q^2)$ are related to the ones of $P_{7,9,10}$ as
\begin{align}\label{eq:C_phenomenological}
C_{7,9,10}(\mu_c)=\frac{4\pi}{\alpha_s(\mu_c) }\left[ V_{cd}^*V_{ud} C_{7,9,10}^{\text{eff}(d)}(\mu_c)+ V_{cs}^*V_{us} C_{7,9,10}^{\text{eff}(s)}(\mu_c) \right] \, .
\end{align}
Using $\sum_{q=d,s,b} V_{cq}^*V_{uq} =0$ makes manifest  that the coefficients are GIM or Cabibbo suppressed, specifically,
$L(m_s^2,q^2)-L(0,q^2)={\cal{O}}(m_s^2/m_c^2)$ at high $q^2$.

\begin{figure}[!htb]
 \centering
 \includegraphics[width=0.8\textwidth]{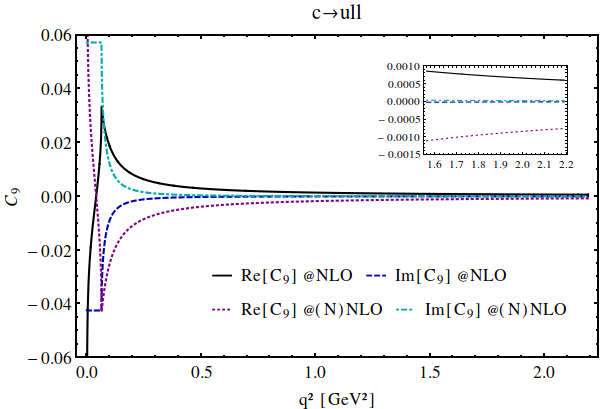}
 \caption{The effective  coefficient $C_{9}(\mu_c=m_c)$ given in Eq.~(\ref{eq:C_phenomenological}) at NLO  and the pure (N)NLO-terms in the SM. 
The two-loop matrix elements of $P_{1,2}$ are not known presently and not included. See text for details. }
 \label{fig:WC}
\end{figure}

The effective coefficient $C_9$,  Eq.~(\ref{eq:C_phenomenological}), in the SM is shown in Fig.~\ref{fig:WC}. $C_7$ is not shown because its $q^2$-dependence is 
negligible.
Note that 
$C_{10}=0$, and that  $C_7$ and $C_9$ 
are primarily set by the matrix elements of $P_{1,2}$.
For $\mu_c=m_c$, $C_{7} \simeq( -0.0011-0.0041 i)$ and $C_9 \simeq  -0.021 X_{ds}$, where $X_{ds}=(V_{cd}^*V_{ud}L(m_d^2,q^2)+V_{cs}^*V_{us}L(m_s^2,q^2))$.
 Varying $m_c/\sqrt2\le\mu_c\le\sqrt2m_c$ we find $(-0.0014-0.0054i)\le C_7\le(-0.00087-0.0033i)$ and 
 $-0.060X_{ds}(\mu_c=\sqrt2m_c)\le C_9\le0.030 X_{ds}(\mu_c=m_c/\sqrt2)$.
 For $q^2 \gtrsim 1 \, \mbox{GeV}^2$, we obtain as a result a small SM contribution, $|C_{9}| \lesssim 5\cdot10^{-4}$.  
 
The  one-loop contribution to $C_9$ is  suppressed by cancellations between  $\tilde C_1$ and $\tilde C_2$. Therefore,
 the two-loop matrix element of $P_{1,2}^{(q)}(\mu_c)$, $q=d,s$ inducing $C_9$,  of the
 order $ |V_{cd}^* V_{ud}| \times \alpha_s(\mu_c)/ (4 \pi) \times $GIM-type $ m_s^2/m_c^2$-suppression at $q^2={\cal{O}}(m_c^2)$\footnote{This behavior  is also supported by a related calculation in $b \to s  ll$ decays \cite{Greub:2008cy}.},
 could numerically be of  similar size as  the (N)NLO one.

\subsection{Phenomenology}\label{sec:pheno}

In this section we study the SM phenomenology of $D^+\to\pi^+\mu^+\mu^-$ decays and introduce SM null tests.
Decay distributions are given  in App.~\ref{app:observables}, and the requisite form factors $f_i$ are defined in App.~\ref{app:form_factors}.
In particular, in our numerical analysis the vector form factor $f_+$ is taken from data \cite{Amhis:2014hma}, and the dipole one $f_T$ is related to $f_+$ through the (improved) Isgur-Wise relations at low and high $q^2$, between which we interpolate,  {\it cf.} App.~\ref{app:form_factors}. A third form factor, $f_0$, does not contribute 
at short distances
as it multiplies $C_{10}$, which vanishes in the SM. $f_+(q^2),f_T(q^2)$ and  $f_0(q^2)$, which can contribute in  SM extensions, can be seen in Fig.~\ref{fig:form_factors}. 
In our calculation we expand squared amplitudes to order $\alpha_s^2$ and apply the pole mass for $m_c$ in matrix elements.

Integrating the distribution in different $q^2$-bins yields the non-resonant SM branching ratios given in Table~\ref{tab:BRSMpi}.
The first uncertainty given corresponds  to the  normalization, which is dominated by the $D$-lifetime, relative to which CKM uncertainties are subdominant.
The dominant theory uncertainty stems from the charm scale $\mu_c$, which here is varied within  $m_c/\sqrt2\le\mu_c\le\sqrt2m_c$.
The effect of a larger upper limit on $\mu_c$  is to enhance (decrease) the branching ratios at low (high) $q^2$. For instance, allowing for values of $\mu_c$ as large as 4 GeV doubles (cuts into halves)
the branching ratios  obtained for $\mu_c=\sqrt{2} m_c$ at low (high) $q^2$. Consequently, the effect on the full $q^2$-range of integration averages out.
The other scales are varied within $m_{W,b}/2 \leq \mu_{W,b} \leq 2 m_{W,b}$.
Uncertainties due to power corrections are not included. 
Electroweak corrections, which are subleading relative to QCD-ones, are neglected.
We checked this explicitly
by calculating the effects of electromagnetic mixing among  the $P_i$ at leading order \cite{Bobeth:2003at,Huber:2005ig}.
Additional uncertainties from $\alpha_s(m_Z)=0.1185 \pm 0.0006$ amount to a few percent.

Further non-resonant SM branching fractions for inclusive $c\to ull$ decays and additional $D\to Pll$ decays  are also worked out and given  in App.~\ref{app:branching_fractions}.
Our findings are consistent with \cite{Fajfer:2002gp,Fajfer:2005ke},
but disagree with those of   \cite{Burdman:2001tf,Paul:2011ar,Wang:2014uiz} by orders of magnitude.
As already discussed around Eq.~(\ref{eq:large_logarithm}), this goes back to  the  inclusion of  light quark masses  in   \cite{Burdman:2001tf,Paul:2011ar,Wang:2014uiz}  in the matching at $\mu_W$.

\begin{table}[!htb]
 \centering
 \begin{tabular}{c|c|c}
  $q^2$-bin                                               &  ${\cal{B}}(D^+\to\pi^+\mu^+\mu^-)^{\rm SM}_{\rm nr} $  & $90 \%$ CL limit \cite{Aaij:2013sua}\\
  \noalign{\hrule height 1pt}
  full $q^2$: $~(2m_\mu)^2\le q^2\le(m_{D^+}-m_{\pi^+})^2~$             &  $3.7\cdot10^{-12}\,(\pm1,\pm3,_{-15}^{+16},\pm1,_{-1}^{+3},_{-1}^{+158},_{-12}^{+16})$  & $7.3 \cdot 10^{-8}$ \\
  \hline
  low $q^2$: $0.250^2\,\text{GeV}^2\le q^2\le0.525^2\,\text{GeV}^2$  &  $7.4\cdot10^{-13}\,(\pm1,\pm4,_{-21}^{+23},_{-11}^{+10},_{-1}^{+10},_{-23}^{+238},_{-5}^{+6})$ &  $2.0 \cdot 10^{-8} $\\
  \hline
 high $q^2$: $~~~~~~~~~~q^2 \ge 1.25^2\,\text{GeV}^2~~~~~~~~~~$                            &  $7.4\cdot10^{-13}\,(\pm1,\pm6,_{-14}^{+15},\pm6,_{-1}^{+0},_{-45}^{+136},_{-20}^{+27})$  & $2.6 \cdot 10^{-8}$\\
 \end{tabular}
 \caption{Non-resonant SM branching fractions of $D^+\to\pi^+\mu^+\mu^-$  decays  normalized to the total width.
Non-negligible uncertainties correspond to (normalization, $m_c$, $m_s$, $\mu_W$, $\mu_b$, $\mu_c$, $f_+$),  respectively, and are given in percent.
 In the last column we give the corresponding experimental  $90 \%$ CL upper limits  \cite{Aaij:2013sua}.}
 \label{tab:BRSMpi}
\end{table}

Next we model the contributions from resonances by using  a  (constant width) Breit-Wigner shape for $C_9 \to C_9^\text R$ for vector 
and $C_P \to C_P^\text R$ for pseudoscalar mesons
\begin{align}
 C_9^\text R&=a_\rho e^{i\delta_\rho} \left(\frac1{q^2-m_\rho^2+im_\rho\Gamma_\rho}-\frac13\frac1{q^2-m_\omega^2+im_\omega\Gamma_\omega}\right)+\frac{a_\phi e^{i\delta_\phi}}{q^2-m_\phi^2+im_\phi\Gamma_\phi}  \, , \nonumber\\
 C_P^\text R &=\frac{a_\eta e^{i\delta_\eta}}{q^2-m_\eta^2+im_\eta\Gamma_\eta}+\frac{a_{\eta'}}{q^2-m_{\eta'}^2+im_{\eta'}\Gamma_{\eta'}}\, \, ,
 \label{eq:BW}
\end{align}
where $\Gamma_M$ denotes the total width of resonance $M=\eta^{(\prime)}, \rho, \omega, \phi$ and we safely neglected the SM's CP-violating effects.
Since the branching fraction of  $D^+\to\pi^+\omega$  decays is not measured yet, and also to reduce the number of parameters, one can use isospin to relate  it to the one of the decay $D^+\to\pi^+\rho$  \cite{Fajfer:2005ke}.
While there are clearly corrections to this ansatz for the $\omega$, these are subdominant relative to the dominant contributions from the $\rho$ due to its large width.

Approximating ${\cal{B}}(D^+ \to \pi^+ M (\to \mu^+ \mu^- )) \simeq  {\cal{B}}(D^+ \to M \pi^+) {\cal{B}}(M \to \mu^+ \mu^-)$ and taking the right-hand side from data 
\cite{Agashe:2014kda} and ${\cal{B}}(\eta^\prime \to \mu^+ \mu^-) \sim {\cal{O}}(10^{-7})$ \cite{Agashe:2014kda,Landsberg:1986fd}, 
we obtain
\begin{align}\label{eq:C9R_parameters}
 &a_\phi=0.24_{-0.06}^{+0.05}\;\text{GeV}^2\,,\quad a_\rho=0.17\pm0.02\;\text{GeV}^2\,,\nonumber\\
 &a_\eta=0.00060_{-0.00005}^{+0.00004}\;\text{GeV}^2\,,\quad a_{\eta'}\sim0.0007\;\text{GeV}^2\,.
\end{align}
We note that   the present experimental  upper limit on ${\cal{B}}(D^+ \to \omega \pi^+)$ yields  $a_\omega\lesssim 0.04$,  somewhat below the isospin prediction, 
$ a_\rho/3$.

\begin{figure}[!htb]
 \centering
 \includegraphics[width=0.8\textwidth]{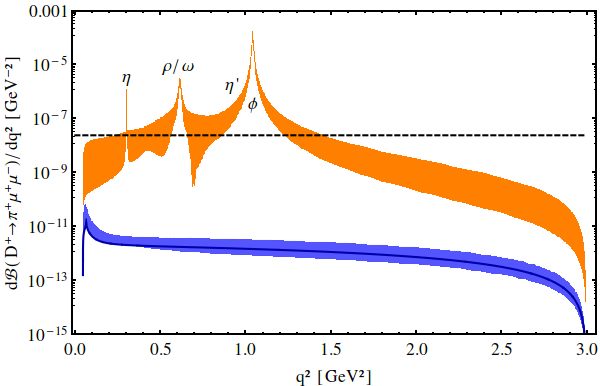}
 \caption{The differential branching fraction $\mathrm d {\cal{B}}(D^+\to\pi^+\mu^+\mu^-)/ \mathrm d q^2$ in the SM.
 The solid blue curve is the non-resonant prediction at $\mu_c=m_c$ and the lighter blue band its $\mu_c$-uncertainty.
 The orange band is the pure resonant contribution taking into account the  uncertainties  specified in Eq.~(\ref{eq:C9R_parameters}) at 1 $\sigma$ and varying the relative strong  phases.
 The dashed black line denotes the 90\% CL experimental upper limit \cite{Aaij:2013sua}.}
 \label{fig:BrSM}
\end{figure}

The SM differential branching fraction  of $D^+\to\pi^+\mu^+\mu^-$ decays is shown in Fig.~\ref{fig:BrSM}. The dominant resonance contributions above the $\phi$-peak are due to the
$\phi$ and the $\rho$. The relative strong phases $\delta_{\phi, \rho ,\eta}$ are varied independently  within $-\pi$ and $\pi$. 
The dominant uncertainty stems from the unknown phases, only near the resonance peaks the uncertainties in the factors $a_M$ become noticeable.
At high $q^2$ the resonances die out with increasing $q^2$, however slowly. For instance, we obtain $|C_9^\text R (1.5  \, \mbox{GeV}^2)|  \lesssim  0.8 $ and $|C_9^\text R (2  \, \mbox{GeV}^2)| \lesssim 0.4$, exceeding by many orders of magnitude the SM short-distance contribution to $Q_9$.

We learn the following: There is room for new physics below the current search limits  \cite{Aaij:2013sua}  and above the resonance contribution; at very high, and very low $q^2$. In either case it will require large BSM contributions to the Wilson coefficients to be above the resonant background.
We will quantify this in Sec.~\ref{sec:BSM}.

The dominance of resonances in the decay rate for SM-like Wilson coefficients is common to all $c \to u l^+l^-$ induced processes, such as
inclusive $D \to X_u l^+l^-$, or other exclusive decays, {\it e.g.},  $D \to \pi \pi l^+l^-$ \cite{Cappiello:2012vg} and  $\Lambda_c \to p l^+l^-$. 
Choosing $c \to u l^+l^-$ induced decay modes other than $D^+ \to \pi^+ l^+ l^-$ does not help gaining BSM sensitivity in the dilepton spectrum, however, other modes may allow to construct more advantageous observables. 
Here we discuss opportunities in semileptonic exclusive decays  with observables where the resonance contribution
is not obstructing SM tests.

Clean SM tests are provided by  the angular distribution in $D \to \pi l^+ l^-$ decays, notably, the lepton forward-backward asymmetry $A_{\rm FB}$ 
and the "flat" term \cite{Bobeth:2007dw}, $F_H$, see App.~\ref{app:observables}. Both observables are null tests of the SM and require  scalar/pseudoscalar operators and tensors to be non-negligible.
A promising avenue to probe operators with Lorentz structures closer  to the ones present in the SM is to study CP-asymmetries in the rate
\begin{align}  \label{eq:ACP}
A_{CP} (q^2) =\frac{\mathrm d\Gamma/\mathrm dq^2-\mathrm d \bar \Gamma/\mathrm dq^2}{ \int_{q^2_\text{min}}^{q^2_\text{max}} \mathrm d q^2 (\mathrm d\Gamma/ \mathrm d q^2 + \mathrm d \bar \Gamma/\mathrm dq^2) } \, ,
\end{align}
where $\mathrm d \bar \Gamma/\mathrm dq^2$ denotes the differential decay rate of the CP-conjugated mode, $D^- \to \pi^- l^+ l^-$.
The difference of the widths can be written as
\begin{align}
 \frac{\mathrm d  \Gamma }{\mathrm dq^2}-\frac{\mathrm d \bar \Gamma}{\mathrm dq^2}
 &=-\frac{G_F^2\alpha_e^2}{384\pi^5m_D^3}\sqrt{\lambda^3(m_D^2,m_P^2,q^2)\left(1-4\frac{m_l^2}{q^2}\right)}\left(1+2\frac{m_l^2}{q^2}\right) \\
 &\times\left( \mathrm{Im}[V_{cd}^*V_{ud} (V_{cs} V_{us}^*)]\mathrm{Im}[c_dc_s^*] +\mathrm{Im}[V_{cd}^*V_{ud} \Delta_9^*] \mathrm{Im}[c_d]f_+ 
        +\mathrm{Im}[ V_{cs}^*V_{us} \Delta_9^*] \mathrm{Im}[c_s]f_+ \right) \,, \nonumber 
\end{align}
where the first term in the last row corresponds to the tiny SM prediction whereas the ones driven by  $\Delta_9=C_9^{BSM}+C_9^\prime$ correspond to possible BSM contributions, and
\begin{align}
 &c_d=\frac{4\pi}{\alpha_s}2C_7^{\text{eff}(d)}f_T\frac{m_c}{m_D}+C_9^\text R|_{\rho-only} \frac{f_+}{V_{cd}^* V_{ud}}\,, \\
 &c_s=\frac{4\pi}{\alpha_s}2C_7^{\text{eff}(s)}f_T\frac{m_c}{m_D} +C_9^\text R|_{\phi-only} \frac{f_+}{V_{cs}^* V_{us}} \, .
\end{align}
Here we neglect all resonances other than the $\phi$ as the latter is dominant on the $\phi$, and the $\rho$, as it is wide. To avoid double-counting we drop the perturbative contributions to
$C_9^{\text{eff}(d,s)}$ in $c_{d,s}$, respectively. The resonance contributions allow to evade the otherwise strong GIM-suppression, a feature already exploited
in probing BSM CP-violation in dipole operators on or near the  $\phi$ resonance \cite{Fajfer:2012nr}.
In the SM CP-violation is tiny due to the smallness of  $\mathrm{Im}[V_{cd}^*V_{ud}(V_{cs} V_{us}^*)]$. We find $|A_{CP}^{\rm SM}(q^2)|<5\cdot10^{-3}$, peaking
at $q^2 \sim m_\phi^2$, where we normalized to the sum of the widths integrated over the full $q^2$-region.
We conclude that while there are large uncertainties related to the phenomenological model for $C_9^\text R$,  it allows to see large BSM effects. We show this explicitly in
Sec.~\ref{sec:BSM}, where we also study LFV decays.

\section{Beyond the Standard Model \label{sec:BSM}}

We discuss testable BSM effects model-independently in Sec.~\ref{sec:MIA} and within leptoquark models, which are introduced briefly in Sec.~\ref{sec:LQ},
in Sec.~\ref{sec:LQ_phenomenology}.

\subsection{Model-independent Analysis \label{sec:MIA}}

To study BSM effects we extend the operator basis (\ref{eq:Q7})-(\ref{eq:Q10})
\begin{align} \label{eq:LeffBSM}
 \mathcal L_\text{eff}^\text{weak}(\mu\sim m_c)=\frac{4G_F}{\sqrt 2}\frac{\alpha_e}{4\pi}\sum_iC_i^{(l)}Q_i^{(l)}\,, \quad \quad \quad (c \to u l^+ l^-) \, ,
\end{align}
where 
\begin{align} \nonumber
 &Q_9^{(l)}=(\bar u\gamma_{\mu}P_Lc)\left(\overline l\gamma^{\mu}l\right)\,,&& Q_9^{(l) \prime}=(\bar u\gamma_{\mu}P_Rc)\left(\overline l\gamma^{\mu}l\right)\,,\label{eq:Q9_phenomenological}\\
  \nonumber
 &Q_{10}^{(l)}=(\bar u\gamma_{\mu}P_Lc)\left(\overline l\gamma^{\mu}\gamma_5l\right)\,,&& Q_{10}^{(l) \prime}=(\bar u\gamma_{\mu}P_Rc)\left(\overline l\gamma^{\mu}\gamma_5l\right)\,,\\
 &Q_S^{(l)}=(\bar uP_Rc)\left(\bar ll\right)\,,&& Q_S^{(l) \prime}=(\bar uP_Lc)\left(\bar ll\right)\,,\\  \nonumber
 &Q_P^{(l)}=(\bar uP_Rc)\left(\bar l\gamma_5l\right)\,,&& Q_P^{(l) \prime}=(\bar uP_Lc)\left(\bar l\gamma_5l\right)\,,\\  \nonumber
 &Q_T^{(l)}=\frac12(\bar u\sigma^{\mu\nu}c)\left(\bar l\sigma_{\mu\nu}l\right)\,, 
&&Q_{T5}^{(l)}=\frac12(\bar u\sigma^{\mu\nu}c)\left(\bar l\sigma_{\mu\nu}\gamma_5l\right)\,.\label{eq:QT5} 
\end{align}
As we use muonic modes frequently, in the following Wilson coefficients and operators without
a lepton flavor index are understood as muonic ones, that is $C_i^{(\mu)} =C_i$ etc. 

Neglecting the SM Wilson coefficients, we find the following constraints on the BSM Wilson coefficients from  the  limits on the branching fraction of $D^+\to\pi^+\mu^+\mu^-$ 
given in Table \ref{tab:BRSMpi}
in the high $q^2$-region ($\sqrt{q^2}\ge1.25\,\text{GeV}$) at CL=90\%  
\begin{align}
 &0.9|C_9+C_9'|^2+0.9|C_{10}+C_{10}'|^2+4.1|C_S+C_S'|^2+4.2|C_P+C_P'|^2+1.1|C_T|^2+1.0|C_{T5}|^2\nonumber\\
 &+0.6\,\mathrm{Re}[(C_9+C_9')C_T^*]+1.2\,\mathrm{Re}[(C_{10}+C_{10}')(C_P+C_P')^*]\nonumber\\
 &+2.3|C_7|^2+2.8\mathrm{Re}[C_7(C_9+C_9')^*]+0.8\,\mathrm{Re}[C_7C_T^*]\lesssim1\,.
\end{align} 
Analogous constraints in the  full $q^2$-region are somewhat stronger. They read
\begin{align}
 &1.3|C_9+C_9'|^2+1.4|C_{10}+C_{10}'|^2+2.2|C_S+C_S'|^2+2.3|C_P+C_P'|^2+0.9|C_T|^2+0.8|C_{T5}|^2\nonumber\\
 &+0.9\,\mathrm{Re}[(C_9+C_9')C_T^*]+1.0\,\mathrm{Re}[(C_{10}+C_{10}')(C_P+C_P')^*]\nonumber\\
 &+3.7|C_7|^2+4.4\mathrm{Re}[C_7(C_9+C_9')^*]+1.3\,\mathrm{Re}[C_7C_T^*]\lesssim1\,.
\end{align} 

The branching fraction $\mathcal B(D^0\to\mu^+\mu^-)<6.2\cdot10^{-9}$  at CL=90\% \cite{Agashe:2014kda} provides complementary  constraints as
\begin{align}
 |C_S-C_S'|^2+|C_P-C_P'+0.1(C_{10}-C_{10}')|^2\lesssim0.007\,.
\end{align}
Thus, $D\to\pi\mu\mu$ is sensitive to the complete set of operators, however, the purely leptonic decays put stronger constraints on scalar and pseudoscalar operators.

\begin{figure}[!htb]
 \centering
 \includegraphics[width=0.8\textwidth]{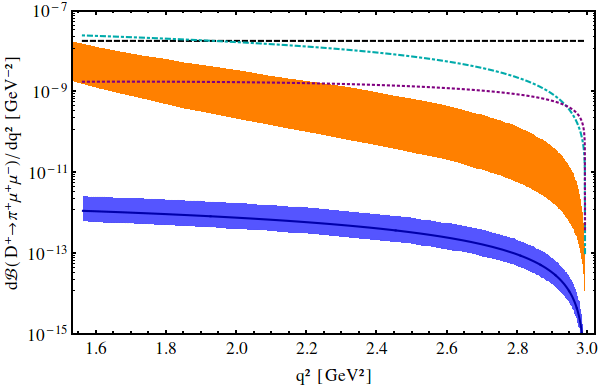}
 \caption{The differential branching fraction $\mathrm d {\cal{B}}(D^+\to\pi^+\mu^+\mu^-)/\mathrm d q^2$ at high $q^2$.
 The solid blue curve is the non-resonant SM prediction at $\mu_c=m_c$ and the lighter blue band its $\mu_c$-uncertainty, the dashed black line denotes the 90\% CL experimental upper limit \cite{Aaij:2013sua} and the orange band shows the resonant contributions.
 The additional curves illustrate two viable, sample BSM scenarios, $|C_9|=|C_{10}|=0.6$ (dot-dashed cyan curve) and $|C_i^{(\prime)}|=0.04$ (dotted purple curve).
 In the latter case all BSM coefficients have been set simultaneously to this value.}
 \label{fig:BSM}
\end{figure}

Barring cancellations, we find, consistent with \cite{Fajfer:2015zea}, 
$|C_{9,10}^{(\prime)}| \lesssim 1$, which can exceed the resonance contribution at high $q^2$.
Assuming no further flavor suppression for the BSM contribution $g^2/\Lambda^2$ (weakly-induced tree level) or
$g^4/ (16  \pi^2 \Lambda^2)$ (weak loop), the limits on $C_{9,10}^{(\prime)}$
  imply  quite mild constraints for the scale of new physics: $\Lambda \gtrsim {\cal{O}}(5)$ TeV or $\Lambda$ around the electroweak scale, respectively.
 With $SU(2)_L$-relations $C_9=-C_{10}$ the bounds on  new physics ease by a factor of $1/\sqrt{2}$.
 Analogous constraints on the other coefficients read $| C_{T,T5}| \lesssim 1$ and $|C_{S,P}^{(\prime)} | \lesssim 0.1$. 
 In Fig.~\ref{fig:BSM} we illustrate BSM effects in the $D^+ \to \pi^+ \mu^+ \mu^- $ differential branching fraction at high $q^2$  
 with two  viable choices for BSM-induced Wilson coefficients. As anticipated, the BSM distributions can exceed the SM one.

 Constraints on $c \to u ee$ modes are weaker than the  $c \to u \mu \mu$ ones, $\mathcal B(D^+\to\pi^+e^+e^-)<1.1\cdot10^{-6}$ and $\mathcal B(D^0\to e^+e^-)<7.9\cdot10^{-8}$  at CL=90\% \cite{Agashe:2014kda}, and imply
 \begin{align} \nonumber
 &\left|C_{S,P}^{(e)}-C_{S,P}^{(e)\prime}\right|\lesssim0.3\,, \\
 &\left|C_{9,10}^{(e)}-C_{9,10}^{(e)\prime}\right|\lesssim4\,,\quad\left|C_{T,T5}^{(e)}\right|\lesssim5\,,\quad\left|C_7\left(C_9^{(e)}-C_9^{(e)\prime}\right)\right|\lesssim2\,.
\end{align}

To discuss LFV we introduce the following effective Lagrangian
\begin{align}
 \mathcal L_\text{eff}^\text{weak}(\mu\sim m_c)=\frac{4G_F}{\sqrt 2}\frac{\alpha_e}{4\pi}\sum_i\left(  K_i^{(e)} O_i^{(e)} + K_i^{(\mu)} O_i^{(\mu)} \right)\,, \quad \quad \quad (c \to u e^\pm \mu^\mp) \, ,
\end{align}
where the $K_i^{(l)}$ denote Wilson coefficients and the operators $O_i^{(l)}$  read
\begin{align}
 &O_9^{(e)}=(\bar u\gamma_{\mu}P_Lc)\left(\overline e \gamma^{\mu} \mu\right)\,,&& O_9^{(\mu)}=(\bar u\gamma_{\mu}P_Lc)\left(\overline \mu \gamma^{\mu} e\right)\,,\label{eq:Z9_phenomenological}
\end{align}
and all others in analogous notation to Eq.~(\ref{eq:Q9_phenomenological}).
The LFV Wilson coefficients  are constrained by $ \mathcal B(D^0\to e^+\mu^-+e^-\mu^+)<2.6\cdot10^{-7}$,
 $ \mathcal B(D^+\to\pi^+e^+\mu^-)<2.9\cdot10^{-6}$ and $\mathcal B(D^+\to\pi^+e^-\mu^+)<3.6\cdot10^{-6}$ 
at CL=90\% \cite{Agashe:2014kda} as
\begin{align} \nonumber 
 &\left|K_{S,P}^{(l)}-K_{S,P}^{(l)\prime}\right|\lesssim0.4\,,\\
 &\left|K_{9,10}^{(l)}-K_{9,10}^{(l)\prime}\right|\lesssim6\,,\quad\left|K_{T,T5}^{(l)}\right|\lesssim7\, , \quad \quad \quad l=e,\mu \, .\label{eq:Ki}
\end{align}

The observables in the $D \to P l^+ l^-$ angular distribution, $A_{\rm FB}$ and $F_H$, Eqs.~(\ref{eq:AFB}), (\ref{eq:FH}) can be sizable while respecting the 
model-independent bounds.
We find that, upon $q^2$-integration,   
$|A_{\rm FB}(D^+\to\pi^+\mu^+\mu^-)|\lesssim 0.6$, $|A_{\rm FB}(D^+\to\pi^+e^+e^-)|\lesssim0.8$, $F_H(D^+\to\pi^+\mu^+\mu^-)\lesssim1.5$ and $F_H(D^+\to\pi^+e^+e^-)\lesssim1.6$, where $F_H^\text{SM}$ is below permille level, allowing to
signal BSM physics. Here,
the resonance contributions have been taken into account in the normalization to the decay rate, and both numerator and denominator  (the decay rate) have been integrated
from  $q^2_\text{min}=1.25^2\,\text{GeV}^2$ to $q^2_\text{max}=(m_{D^+}-m_{\pi^+})^2$.
As the LFV bounds (\ref{eq:Ki}) are even weaker than the ones on the dielectron modes, sizable contributions to LFV angular observables are allowed as well. 
Knowing the size of LFV in
more than one observable would allow to pin down the operator structure and provide clues about the underlying model.

\subsection{\texorpdfstring{$c\to ull$}{ctoull} Generating Models   \label{sec:LQ}}

Several models generating $c\to ull$ transitions were studied, for instance,
Little Higgs models   \cite{Fajfer:2005ke,Paul:2011ar},
Minimal Supersymmetric Standard Models   \cite{Fajfer:2001sa,Burdman:2001tf,Fajfer:2007dy,Wang:2014uiz},
two Higgs doublet models \cite{Fajfer:2001sa}, an up vector-like quark singlet \cite{Fajfer:2007dy} 
and models with warped extra dimensions \cite{Delaunay:2012cz,Paul:2012ab}.
In all models except for the supersymmetric ones  the $D \to \pi l^+ l^-$ branching fraction is found to be less than the resonance contributions.
In the supersymmetric models the branching ratio can be  close to the experimental limits.
Non-vanishing asymmetries  that could be $A_{\rm FB}$, $A_{CP}$ and the CP-asymmetry of $A_{\rm FB}$ are generically  induced in BSM models
 \cite{Fajfer:2005ke,Paul:2011ar,Burdman:2001tf,Wang:2014uiz, Delaunay:2012cz,Paul:2012ab}.

Here we study effects of leptoquarks generating $c\to ull$ transitions in a bottom-up approach. We note that in Grand Unified Theories further
model building for some representations is required \cite{Dorsner:2012nq}.
For renormalizable up-type scalar and vector $SU(2)_L$ singlet, doublet and triplet leptoquarks within the SM ($SU(3)_C$, $SU(2)_L$, $U(1)_Y$) gauge group \cite{Buchmuller:1986zs} we find after Fierzing the effective contributions shown in Table~\ref{tab:cull_leptoquarks}. 
Baryon number and lepton number are conserved in the interactions. Note that models $S_1$ and $S_2$ contain two couplings each.
Leptoquark effects in $S_1$ have been  discussed in  \cite{Fajfer:2008tm}.

\begin{table}[!htb]
 \centering
 \begin{tabular}{c|c|c}
  $\subset\mathcal L_{LQ}$                    &                   $(SU(3)_C, SU(2)_L ,Y)$                                                        &  effective vertices  \\
  \noalign{\hrule height 1pt}
  $\left(\lambda_{S_1L}\mathbf Q_L^Ti\tau_2\mathbf L_L+\lambda_{S_1R}q_Rl_R\right)S_1^\dagger\quad$ & $(3,1,-1/3)$           &  $-\frac{\left(\lambda_{S_1R}^{(ql')}\right)^*\lambda_{S_1R}^{(q'l)}}{2m_{S_1}^2}\left(\bar q_R\gamma_\mu q_R'\right)\left(\bar l_R\gamma^\mu l_R'\right)$  \\
                                                                                                             &           &  $-\frac{\left(\lambda_{S_1L}^{(ql')}\right)^*\lambda_{S_1L}^{(q'l)}}{2m_{S_1}^2}\left(\bar q_L\gamma_\mu q_L'\right)\left(\bar l_L\gamma^\mu l_L'\right)$  \\
                                                                                                               &         &  $-\frac{\left(\lambda_{S_1R}^{(ql')}\right)^*\lambda_{S_1L}^{(q'l)}}{2m_{S_1}^2}\left(\bar q_Lq_R'\right)\left(\bar l_Ll_R'\right)$  \\
                                                                                                              &          &  $-\frac{\left(\lambda_{S_1L}^{(ql')}\right)^*\lambda_{S_1R}^{(q'l)}}{2m_{S_1}^2}\left(\bar q_Rq_L'\right)\left(\bar l_Rl_L'\right)$  \\
                                                                                                               &         &  $-\frac{\left(\lambda_{S_1R}^{(ql')}\right)^*\lambda_{S_1L}^{(q'l)}}{8m_{S_1}^2}\left(\bar q\sigma_{\mu\nu} q'\right)\left(\bar l_L\sigma^{\mu\nu}l_R'\right)$  \\
                                                                                                              &          &  $-\frac{\left(\lambda_{S_1L}^{(ql')}\right)^*\lambda_{S_1R}^{(q'l)}}{8m_{S_1}^2}\left(\bar q\sigma_{\mu\nu} q'\right)\left(\bar l_R\sigma^{\mu\nu}l_L'\right)$  \\
  \hline
  $\left(\lambda_{S_2L}\bar q_R\mathbf L_L+\lambda_{S_2R}\bar{\mathbf Q}_Li\tau_2l_R\right)S_2^\dagger\quad $ & $(3,2,-7/6)$  &  $-\frac{\lambda_{S_2R}^{(ql')}\left(\lambda_{S_2R}^{(q'l)}\right)^*}{2m_{S_2}^2}(\bar q_L\gamma_\mu q_L')\left(\bar l_R\gamma^\mu l_R'\right)$  \\
                                                                                                                &        &  $-\frac{\lambda_{S_2L}^{(ql')}\left(\lambda_{S_2L}^{(q'l)}\right)^*}{2m_{S_2}^2}(\bar q_R\gamma_\mu q_R')\left(\bar l_L\gamma^\mu l_L'\right)$  \\
                                                                                                                &        &  $-\frac{\lambda_{S_2R}^{(ql')}\left(\lambda_{S_2L}^{(q'l)}\right)^*}{2m_{S_2}^2}(\bar q_Lq_R')\left(\bar l_Ll_R'\right)$  \\
                                                                                                                &        &  $-\frac{\lambda_{S_2L}^{(ql')}\left(\lambda_{S_2R}^{(q'l)}\right)^*}{2m_{S_2}^2}(\bar q_Rq_L')\left(\bar l_Rl_L'\right)$  \\
                                                                                                               &         &  $-\frac{\lambda_{S_2R}^{(ql')}\left(\lambda_{S_2L}^{(q'l)}\right)^*}{8m_{S_2}^2}\left(\bar q\sigma_{\mu\nu} q'\right)\left(\bar l_L\sigma^{\mu\nu}l_R'\right)$  \\
                                                                                                                 &       &  $-\frac{\lambda_{S_2L}^{(ql')}\left(\lambda_{S_2R}^{(q'l)}\right)^*}{8m_{S_2}^2}\left(\bar q\sigma_{\mu\nu} q'\right)\left(\bar l_R\sigma^{\mu\nu}l_L'\right)$  \\
  \hline
  $\left(\lambda_{S_3}\mathbf Q_L^Ti\tau_2\vec\tau\mathbf L_L\right)\cdot\vec{S_3}^\dagger\quad$ & $ (3,3,-1/3)$              &  $-\frac{\left(\lambda_{S_3}^{(ql')}\right)^*\lambda_{S_3}^{(q'l)}}{2m_{S_3}^2}\left(\bar q_L\gamma_\mu q_L'\right)\left(\bar l_L\gamma^\mu l_L'\right)$  \\
  \hline
  $\lambda_{\tilde V_1}\bar q_R\gamma_\mu l_R\left(\tilde V_1^\mu\right)^\dagger\quad$ & $ (3,1,-5/3)$                        &  $\frac{\lambda_{\tilde V_1}^{(ql')}\left(\lambda_{\tilde V_1}^{(q'l)}\right)^*}{m_{\tilde V_1}^2}(\bar q_R\gamma_\mu q_R')\left(\bar l_R\gamma^\mu l_R'\right)$  \\
  \hline
  $\lambda_{V_2}\mathbf Q_L\gamma_\mu l_R\left(V_2^\mu\right)^\dagger\quad $ & $ (3,2,-5/6)$                                   &  $\frac{\left(\lambda_{V_2}^{(ql')}\right)^*\lambda_{V_2}^{(q'l)}}{m_{V_2}^2}\left(\bar q_L\gamma_\mu q_L'\right)\left(\bar l_R\gamma^\mu l_R'\right)$  \\
  \hline
  $\lambda_{\tilde V_2}q_R\gamma_\mu\mathbf L_L\left(\tilde V_2^\mu\right)^\dagger\quad $ & $ (3,2,1/6)$                      &  $\frac{\left(\lambda_{\tilde V_2}^{(ql')}\right)^*\lambda_{\tilde V_2}^{(q'l)}}{m_{\tilde V_2}^2}\left(\bar q_R\gamma_\mu q_R'\right)\left(\bar l_L\gamma^\mu l_L'\right)$  \\
  \hline
  $\lambda_{V_3}\bar{\mathbf Q}_L\gamma_\mu\vec\tau\mathbf L_L\cdot\left(\vec{V_3}^\mu\right)^\dagger\quad $&$ (3,3,-2/3)$   &  $\frac{2\lambda_{V_3}^{(ql')}\left(\lambda_{V_3}^{(q'l)}\right)^*}{m_{V_3}^2}(\bar q_L\gamma_\mu q_L')\left(\bar l_L\gamma^\mu l_L'\right)$  \\
 \end{tabular}
 \caption{Leptoquark-fermion interactions, quantum numbers, with hypercharge  $Y=Q_e-T_3$, and effective $c\to u(l')^+l^-$ vertices via Fierz identities. $\tau_a, a=1,2,3$ denote the Pauli matrices,  and $\vec \tau \cdot \vec X=\sum _a \tau_a X_a$ for $X=S_3,V_3$. SM  $SU(2)_L$-doublets are $\mathbf Q(3,2,1/6)$ and $\mathbf L(1,2,-1/2)$, 
 $q,q' =u,c$, and $l,l'=e,\mu$.}
 \label{tab:cull_leptoquarks}
\end{table}

We uniformly denote by $M$ the mass of the leptoquarks but note that they differ in general
depending on the representation. We assume degenerate $SU(2)_L$-plet masses to comply with the constraints from oblique parameters \cite{Davidson:2010uu}.
Our effective vertices  agree with and extend those in \cite{Davidson:1993qk} by considering tensor operators and relative signs.
The  Wilson coefficients induced by tree-level leptoquark exchanges read as
\begin{align}\label{eq:LQWilson}
 &C_{9,10}=\frac{\sqrt2\pi}{G_F\alpha_e}k_{9,10}\frac{\lambda_i^I\left(\lambda_i^J\right)^*}{M^2}\,,&&C_{9,10}'=\frac{\sqrt2\pi}{G_F\alpha_e}k_{9,10}'\frac{\lambda_j^I\left(\lambda_j^J\right)^*}{M^2}\,,\\  \nonumber 
 &C_S=C_P=\frac{\sqrt2\pi}{G_F\alpha_e}k_{S,P}\frac{\lambda_i^I\left(\lambda_j^J\right)^*}{M^2}\,,&&C_S'=-C_P'=\frac{\sqrt2\pi}{G_F\alpha_e}k_{S,P}'\frac{\lambda_j^I\left(\lambda_i^J\right)^*}{M^2}\,,\\  \nonumber 
 &C_T=\frac{\sqrt2\pi}{G_F\alpha_e}k_T\left(\frac{\lambda_i^I\left(\lambda_j^J\right)^*}{M^2}+\frac{\lambda_j^I\left(\lambda_i^J\right)^*}{M^2}\right) \,,&&C_{T5}=\frac{\sqrt2\pi}{G_F\alpha_e}k_{T5}\left(\frac{\lambda_i^I\left(\lambda_j^J\right)^*}{M^2}-\frac{\lambda_j^I\left(\lambda_i^J\right)^*}{M^2}\right) \, ,
\end{align}
where  $i,j=L,R$; such indices are nontrivial for scenarios $S_1$ and $S_2$ only, which have two different couplings $\lambda_{L,R}$ each.
The correct values of $i,j$ can also  be read off from Table \ref{tab:cull_leptoquarks}.
The coefficients $k_x$ are given in Table~\ref{tab:coeff}.

\begin{table}[!htb]
 \centering
 \begin{tabular}{|c|c|c|c|c|c|c|c|c|c|c|c|c|c|c|}
 \hline
                &  $I$     &  $J$     &  $i$   &  $j$   &  $k_9$       &  $k_{10}$    &  $k_9'$      &  $k_{10}'$   &  $k_{S,P}$   &  $k_{S,P}'$  &  $k_T$       &  $k_{T5}$  \\
  \noalign{\hrule height 1pt}
  $S_1$         &  $(cl)$  &  $(ul)$  &  $L$   &  $R$   &  $-\frac14$  &  $\frac14$   &  $-\frac14$  &  $-\frac14$  &  $-\frac14$  &  $-\frac14$  &  $-\frac18$  &  $-\frac18$  \\
  \hline
  $S_2$         &  $(ul)$  &  $(cl)$  &  $R$   &  $L$   &  $-\frac14$  &  $-\frac14$  &  $-\frac14$  &  $\frac14$   &  $-\frac14$  &  $-\frac14$  &  $-\frac18$  &  $-\frac18$  \\
  \hline
  $S_3$         &  $(cl)$  &  $(ul)$  &  $L$   &  --  &  $-\frac14$  &  $\frac14$   &  $0$         &  $0$         &  $0$         &  $0$         &  $0$         &  $0$  \\
  \hline
  $\tilde V_1$  &  $(ul)$  &  $(cl)$  &  --  &  $R$   &  $0$         &  $0$         &  $\frac12$   &  $\frac12$   &  $0$         &  $0$         &  $0$         &  $0$  \\
  \hline
  $V_2$         &  $(cl)$  &  $(ul)$  &  $R$   &  --  &  $\frac12$   &  $\frac12$   &  $0$         &  $0$         &  $0$         &  $0$         &  $0$         &  $0$  \\
  \hline
  $\tilde V_2$  &  $(cl)$  &  $(ul)$  &  --  &  $L$   &  $0$         &  $0$         &  $\frac12$   &  $-\frac12$  &  $0$         &  $0$         &  $0$         &  $0$  \\
  \hline
  $V_3$         &  $(ul)$  &  $(cl)$  &  $L$   &  --  &  $1$         &  $-1$        &  $0$         &  $0$         &  $0$         &  $0$         &  $0$         &  $0$  \\
  \hline
 \end{tabular}
 \caption{Coefficient matrix for the leptoquark Wilson coefficients (\ref{eq:LQWilson}) inducing $c \to u ll$.}
 \label{tab:coeff}
\end{table}

\subsection{Leptoquark Phenomenology}\label{sec:LQ_phenomenology}

Experimental constraints on leptoquark couplings are worked out in App.~\ref{app:LQ_constraints}. While generically
$|\lambda^{(ql)}|\lesssim\mathcal O(0.1)\,[M/\text{TeV}]$ for any coupling to the first two generations and for any scenario of Table \ref{tab:cull_leptoquarks},
several  flavor-combinations are more severely  constrained. In addition, bounds for specific models making use of correlations can be much stronger.

The  $|\Delta C|=|\Delta U|=1$ couplings in leptoquark scenarios involving doublet-quarks $\mathbf{Q}$  are subject to constraints from the kaon sector (Table~\ref{tab:LQ_Kaon_constraints}). Corresponding limits  on the Wilson coefficients  for $c \to u l l^{(\prime)}$ are given in Table~\ref{tab:LQ_Kaon_Wilson_coefficients}.
Only the scenarios $\tilde V_1$ and $\tilde V_2$, as well as the $S_1|_R$ and $S_2|_L$ couplings  do not receive  such constraints, hence allow in general for larger effects for $c \to u ll$,
however, decouple without further input from the $K$- and $B$-sector.

\begin{table}[!htb]
 \centering
 \begin{tabular}{c|c|c|c}
                            &  $(ee)$                   &  $(e\mu), (\mu e)$                 &  $(\mu\mu)$  \\
  \noalign{\hrule height 1pt}
  $S_1|_L$                  &  $\lesssim4\cdot10^{-3}$  &  $\lesssim4\cdot10^{-3}$  &  $\lesssim4\cdot10^{-3}$  \\
  $S_2|_R$, $V_2$           &  $\lesssim3\cdot10^{-2}$  &  $\lesssim2\cdot10^{-4}$  &  $\lesssim4\cdot10^{-3}$  \\
  $S_3$, $V_3$              &  $\lesssim4\cdot10^{-3}$  &  $\lesssim2\cdot10^{-4}$  &  $\lesssim4\cdot10^{-3}$  \\
 \end{tabular}
 \caption{Upper limits on the $c\to u ll^{(\prime)}$ Wilson coefficients $|C_{9,10}^{(\prime)}|$ abbreviated as $(l l^{(\prime)})$ in leptoquark scenarios from  kaon decays.
 For $S_{1,2}$ the limits apply to the indicated handedness of couplings only. }
  \label{tab:LQ_Kaon_Wilson_coefficients}
\end{table}

Products of two Wilson coefficients are  constrained by the strong limits on $\mu -e$ conversion and $\mu \to e \gamma$, see Table~\ref{tab:LQ_products_Wilson_coefficients}.
 Future experiments on $\mu\to e\gamma$ \cite{Baldini:2013ke} and $\mu-e$ conversion \cite{Cui:2009zz,Bartoszek:2014mya} can improve the limits by at least two orders of magnitude. 
\begin{table}[!htb]
 \centering
 \begin{tabular}{c|c}
                                                   &  $(ee)(\mu\mu)$, $(e\mu)(\mu e)$  \\
  \noalign{\hrule height 1pt}
  $S_1$                                            &  $\lesssim2\cdot10^{-7}$  \\
  $S_2$                                            &  $\lesssim8\cdot10^{-8}$  \\
  $S_3$, $\tilde V_1$, $V_2$, $\tilde V_2$, $V_3$  &  $\lesssim6\cdot10^{-8}$  \\
  $S_1|_{LR}$, $S_2|_{LR}$                         &  $\lesssim2\cdot10^{-10}$  \\
 \end{tabular}
 \caption{Upper limits on the products of two $c\to ull^{(\prime)}$ Wilson coefficients $|C_i^{(e) (\prime)}\,C_i^{(\mu) (\prime)}|$, "$(ee)(\mu\mu)$", and $|K_i^{(e)(\prime)} K_i^{(\mu)(\prime)}|$, "$(e\mu)(\mu e)$",
 from  $\mu-e$ conversion and $\mu \to e\gamma$.
 The LR-mixing constraints in scenarios $S_{1,2}$ are stronger than the unmixed ones and are given in the last row.}
 \label{tab:LQ_products_Wilson_coefficients}
\end{table}

Further bounds and correlations depend on the flavor structure. To make progress here we study benchmark patterns of leptoquark coupling matrices $\lambda$ put forward in \cite{Varzielas:2015iva} for quark-$\mathbf{L}$-type Yukawa couplings based on flavor symmetries. 
Rows label quark flavors and columns label lepton flavors. 
The use of discrete non-abelian symmetries for the leptons, specifically $A_4$ \cite{deMedeirosVarzielas:2005ax,Altarelli:2005yx}, results in  textures
with "ones" and "zeros", very different  from hierarchical ones in Froggatt-Nielsen $U(1)$-models \cite{Froggatt:1978nt}.
In this work we are mainly concerned with the first two generations, 
so our terminology  reflects features of the  upper left two-by-two submatrix of $\lambda$. With the exception of $D^0 \to \tau e$ and $c \to u \nu \bar \nu$, the third ($\tau, \nu_\tau$) column  is irrelevant to our study.
Similar statements hold for the third ($t,b$) row, which is relevant to $B$-physics, and is linked to charm physics via flavor.
We define

i) a hierarchical flavor pattern with suppression factors for electrons, $\kappa$, and first and second generation quarks,  $\rho_d$ and $\rho$, respectively,
\begin{align}\label{eq:lambda_rhokappa}
\lambda_i\sim
\left(
\begin{array}{ccc}
 \rho_d\kappa  &  \rho_d  &  \rho_d  \\
 \rho\kappa    &  \rho    &  \rho  \\
 \kappa        &  1       &  1  \\
\end{array}
\right)\,,
\end{align}
ii) a single lepton pattern with  negligible electron couplings 
\begin{align}\label{eq:lambda_no-e}
\lambda_{ii} \sim\
\left(
\begin{array}{ccc}
 0  &     *  &  0 \\
 0    &   *    &  0  \\
 0       &  *       &  0  \\
\end{array}
\right)\,,
\end{align}
and iii) a (first two) generation-diagonal "skewed" pattern, that is, $\lambda^{(u\mu)}$ and $\lambda^{(ce)}$ are negligible
\begin{align}\label{eq:skewed}
\lambda_{iii} \sim\
\left(
\begin{array}{ccc}
 *  &     0  &  0 \\
 0    &   *    &  0 \\
 0       &  *       &  0  \\
\end{array}
\right)\,.
\end{align}
The patterns i) and ii) have been explicitly obtained in models where quarks are $A_4$-singlets, hence apply to all $u_R, d_R$ and $\mathbf{Q}$ fields coupling to 
lepton doublets.\footnote{We thank Ivo de Medeiros Varzielas for confirmation.} 
Extension of  \cite{Varzielas:2015iva}  to include
lepton singlets as well as the skewed patterns  iii) and iv), the latter defined in  Eq.~(\ref{eq:skewed2}), is not as straightforward  and requires further model building, which is beyond the scope of this work.  (Note that skewed patterns have been obtained by assigning different quark generations to  different $A_4$ singlet representations \cite{Varzielas:2015iva}.)

Upper limits on rare charm branching fractions for different flavor patterns are given in Table~\ref{tab:LQ_branching_fractions}. Here,
for patterns ii) and iii) we distinguish between leptoquark scenarios  which can escape  kaon bounds, $S_{1,2}$, $\tilde V_{1,2}$,   ii.1) and  iii.1), and  those subject to kaon bounds, $S_3$, $V_{2,3}$,  ii.2) and  iii.2).
If $\kappa$ is small the hierarchical flavor pattern i) effectively  reduces to pattern ii).
	
The $c \to u e^+ e^-$ Wilson coefficients  vanish in patterns ii) and iii). In pattern i) they are driven by $\rho_d \rho \kappa^2$, and correlated with LFV, hence subject to the
bounds in Table~\ref{tab:LQ_products_Wilson_coefficients}.
We find that no BSM signal can be seen in $c \to u e^+ e^-$ branching ratios.

In pattern  ii.1) the muon Wilson coefficients are constrained by $D^+\to\pi^+\mu^+\mu^-$ and $D^0\to\mu^+\mu^-$ as discussed in 
Sec.~\ref{sec:MIA}. For  ii.2) the constraints on the muon Wilson coefficients  are given in Table~\ref{tab:LQ_Kaon_Wilson_coefficients}.
In case of iii) the $c \to u \mu^+ \mu^-$  Wilson coefficients vanish. 

 The dineutrino mode is induced in $S_{2,3}$, $\tilde V_2$ and $V_3$ models because those contain the requisite electromagnetic charge $+2/3e$ leptoquark.
The decay $D \to \pi \nu \bar \nu$ has backgrounds from $D \to \tau (\to \pi \nu) \bar \nu $, which can be controlled by kinematic cuts $q^2>(m_\tau^2-m_{\pi^+}^2)(m_{D^+}^2-m_\tau^2)/m_\tau^2\simeq0.34\,\text{GeV}^2$   \cite{Burdman:2001tf,Kamenik:2009kc}.

The LFV  transition $c \to u \mu^- e^+$ ($c \to u \mu^+ e^-$) is mediated by a generation-diagonally coupling leptoquark with electric charge $5/3 e$ ($-1/3 e$).  
Therefore,
for case iii) either, for charge $-1/3e$,  $\mathcal B(D^0\to\mu^- e^+)$  and  $\mathcal B(\bar D^0\to\mu^+ e^-)$ vanish, or, for charge $5/3e$, $\mathcal B(D^0\to\mu^+ e^-)$ and $\mathcal B(\bar D^0\to\mu^- e^+)$ vanish. Analogous statements hold for $D^+ \to \pi^+ e^\pm \mu^\mp$ decays.
For  iii.1) the LFV Wilson coefficients are $\lesssim\mathcal O(1-10)$, see Eq.~(\ref{eq:Ki}), and
for   iii.2) the constraints on the LFV Wilson coefficients from Table~\ref{tab:LQ_Kaon_Wilson_coefficients} apply.
For  ii) the LFV Wilson coefficients vanish.

\begin{table}[!htb]
 \centering
 \begin{tabular}{c|c|c|c|c|c}
     &  $\mathcal B(D^+\to\pi^+\mu^+\mu^-)$       &  $\mathcal B (D^0\to\mu^+\mu^-)$  &  $\mathcal B (D^+\to\pi^+e^\pm\mu^\mp)$  &  $\mathcal B (D^0\to\mu^\pm e^\mp)$  &  $\mathcal B (D^+\to\pi^+\nu\bar\nu)$  \\
  \noalign{\hrule height 1pt}
  i)   &  SM-like                                    &  SM-like                                                                                   &  $\lesssim2\cdot10^{-13}$                &  $\lesssim7\cdot10^{-15}$            &  $\lesssim3\cdot10^{-13}$  \\
  \hline
 ii.1)  &  $\lesssim7\cdot10^{-8}$ ($2\cdot10^{-8}$)                                                         &  $\lesssim3\cdot10^{-9}$          &  $0$                                 &  $0$                             &  $\lesssim8\cdot10^{-8}$  \\
 ii.2)  &  SM-like                                                                                             &  $\lesssim4\cdot10^{-13}$         &  $0$                                 &  $0$                             &  $\lesssim4\cdot10^{-12}$  \\
  \hline
 iii.1)  &  SM-like                                                                                       &  SM-like                         &  $\lesssim2\cdot10^{-6}$                 &  $\lesssim4\cdot10^{-8}$             &  $\lesssim 2 \cdot 10^{-6}$  \\
  iii.2)  &  SM-like                                    &  SM-like                          &  $\lesssim8\cdot10^{-15}$                &  $\lesssim2\cdot10^{-16}$            &  $\lesssim9\cdot10^{-15}$  \\
 \end{tabular}
 \caption{Branching fractions for  the full $q^2$-region (high $q^2$-region) for different classes of leptoquark couplings, see text.
     Summation of neutrino flavors is understood. 
 "SM-like" denotes a branching ratio which is dominated by resonances or is of similar size  as the resonance-induced one.
 All  $c \to u e^+ e^-$ branching ratios are "SM-like" in the models considered. Note that in the SM $\mathcal B(D^0 \to \mu \mu) \sim 10^{-13}$ \cite{Burdman:2001tf}.}
   \label{tab:LQ_branching_fractions}
\end{table}

\begin{figure}[!htb]
 \centering
 \includegraphics[width=0.48\textwidth]{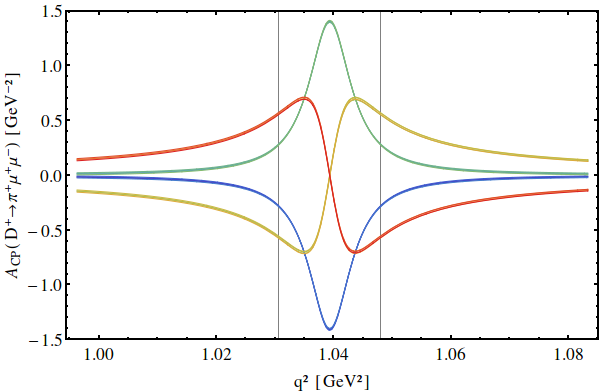}
 \hspace{1em}
 \includegraphics[width=0.48\textwidth]{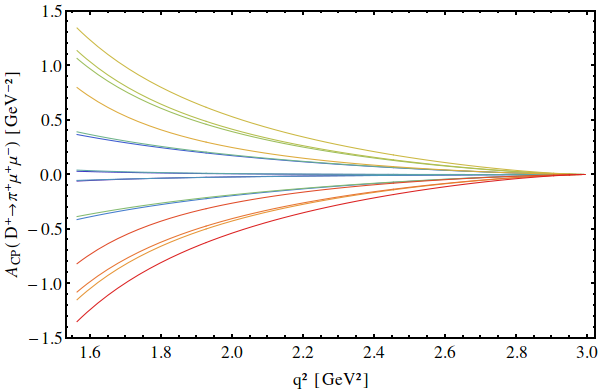}
 \\[2em]
 \includegraphics[width=0.48\textwidth]{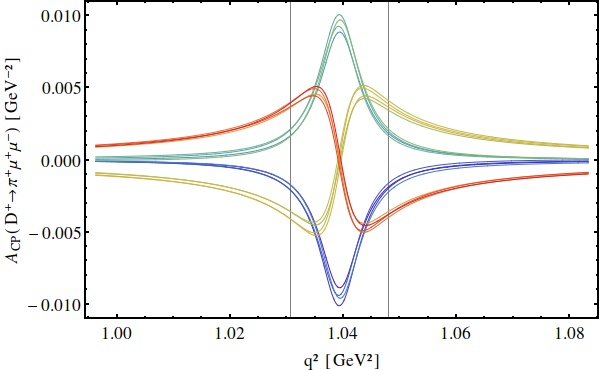}
 \hspace{1em}
 \includegraphics[width=0.48\textwidth]{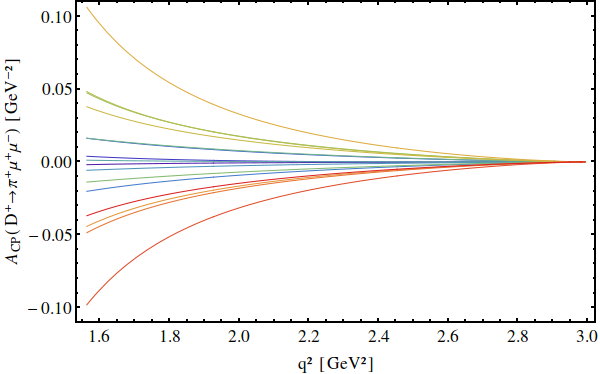}
 \caption{The direct CP-asymmetry $A_{CP}(D^+ \to \pi^+ \mu^+ \mu^-)$ normalized to $[q^2_\text{min}=(m_\phi-5\Gamma_\phi)^2$, $q^2_\text{max}=(m_\phi+5\Gamma_\phi)^2]$ (left plots) and $[q^2_\text{min}=1.25^2\,\text{GeV}^2$, $q^2_\text{max}=(m_{D^+}-m_{\pi^+})^2]$ (right plots) in case of  ii.1) (upper plots) and ii.2) (lower plots) for independent relative strong phases $\delta_{\rho,\phi}\in\{0,\pi/2,\pi,3\pi/2\}$.
 From yellow (upper curves above $\phi$) to red (lower curves above $\phi$) each bunch represents $\delta_\phi=\pi/2,\pi,0,3/2\pi$. }
 \label{fig:ACP}
\end{figure}

Complex couplings are additionally constrained by  electron and neutron electric dipole moments as
$\mathrm{Im}[C_i^{(e)}]\lesssim4\cdot10^{-9}$ and
$\mathrm{Im}[C_i^{(\mu)}]\lesssim4\cdot10^{-6}$,  $i=S,P,T,T5$, respectively.
The  $D^+ \to \pi^+ \mu^+ \mu^-$ CP-asymmetry in the rate, Eq.~(\ref{eq:ACP}), is shown for the muons-only pattern  ii) in Fig.~\ref{fig:ACP}.
Around the  $\phi$-resonance (left-handed plots), $A_{CP}$ scales with the BSM  coefficient $\Delta_9$, as the normalization is driven by the resonances, $C_9^R$, for any BSM coefficient.
At high $q^2$ (right-handed plots) the normalization depends on the value of $\Delta_9$. In the  plot to the upper right  the normalization  is set by $\Delta_9$, 
hence $A_{CP}$ becomes inversely proportional to $\Delta_9$. In the 
plot to the lower right, 
corresponding to a scenario with smaller BSM effects,  ii.2), the normalization is again dominated by the resonances.
Despite the constrained Wilson coefficients
the CP-asymmetry can be sizable around the $\phi$ and above in the high $q^2$-region, in which $|A_{CP}|$ drops towards the endpoint.
If  measured around the $\phi$, a sizable CP-asymmetry, while assuming different values,  can arise  independent of the strong phases. 
For the hierarchical pattern i) $|A_{CP}|$ is $\lesssim 0.003$ on the $\phi$-resonance and $\lesssim 0.03$ at high $q^2$.

Interestingly, there exists an opportunity to also study $\tau$-lepton  couplings in charm,  with $D^0\to\tau^\pm  e^\mp$ decays. The corresponding branching fractions can be inferred from Eq.~(\ref{eq:B_D0mue}); the phase space suppression relative to $D^0\to\mu^\pm  e^\mp$ is about
$8 \cdot 10^{-3}$.  We find, approximately,
\begin{align}
 \mathcal B(D^0\to\tau^\pm e^\mp)&\simeq5\cdot10^{-9}\bigg(\left|1.3\left(K_S^{(e,\tau)}-K_S^{(e,\tau)\prime}\right)\mp\left(K_9^{(e,\tau)}-K_9^{(e,\tau)\prime}\right)\right|^2\nonumber\\
 &+\left|1.3\left(K_P^{(e,\tau)}-K_P^{(e,\tau)\prime}\right)+\left(K_{10}^{(e,\tau)}-K_{10}^{(e,\tau)\prime}\right)\right|^2\bigg)\,.
\end{align}
The limits on the decays $\tau\to e\gamma$ and $\tau\to eee$ are not competitive with those involving muons, however,
$SU(2)$-relations imply constraints on LFV. For (axial)vector couplings  they read  $|K_{9,10}^{(e,\tau)( \prime)}|\lesssim 0.2$ ($ \mathcal B(\tau\to eK)$), significantly weaker than $|K_{9,10}^{(e,\tau)}|\ \lesssim4\cdot10^{-3}$ ($\mathcal B(K^+\to\pi^+\bar\nu\nu)$) and  for (pseudo)scalar Wilson coefficients $|K_{S,P}^{(e,\tau)(\prime)}|\lesssim7\cdot10^{-3}$ ($\mathcal B(K^+\to \bar e\nu)$) \cite{Carpentier:2010ue}.
The hierarchical flavor pattern yields $\mathcal B (D^0\to\tau^\pm e^\mp) \lesssim7\cdot10^{-15}$, while the others,  ii) and iii), give vanishing rates.
 One flavor pattern in which the $SU(2)$-related constraints are absent and which can signal LFV BSM $D^0\to\tau^\pm e^\mp$ decays, is
another skewed one, inspired by \cite{Varzielas:2015iva},
\begin{align}\label{eq:skewed2}
\lambda_{iv} \sim\
\left(
\begin{array}{ccc}
 0 &     0  &  * \\
 *    &   0    &  0 \\
 *       &  0       &  0  \\
\end{array}
\right)\,.
\end{align}
This pattern results in SM-like lepton-diagonal $c \to u l^+ l^-$, $l=e,\mu$ and vanishing flavor off-diagonal $c \to u e^\pm \mu^\mp$ modes, while
${\cal{B}}(D \to \pi \nu \bar \nu)$  can exceed the upper limits given in Table \ref{tab:LQ_branching_fractions}.
Other flavor patterns  result in a different phenomenology, hence, if measured, this allows to learn about
flavor.

\section{Summary \label{sec:summary}}

Rare charm decays into leptons offer genuine avenues to search for BSM physics despite notorious resonance backgrounds.
Semileptonic branching ratios $D \to \pi l^+ l^-$ can signal BSM physics above the $\phi$-resonance  right around the current experimental limit for large BSM contributions, 
see Fig.~\ref{fig:BSM}.
CP-asymmetries, assisted by the resonances, observables in the angular distribution, dineutrino modes  and LFV ones can signal BSM physics for much smaller BSM contributions, because those correspond to SM null tests.
Model-independent constraints  are given in Sec.~\ref{sec:MIA}.

We work out  correlations in several flavor benchmarks for scalar and vector leptoquark scenarios  that induce $c \to u l l^{(\prime)}$ modes.
The main results on the leptoquark phenomenology are given in Sec.~\ref{sec:LQ_phenomenology}.
We find that hierarchical flavor patterns such as (\ref{eq:lambda_rhokappa}) allow only for rather limited effects in charm due to the correlations with
other sectors which are subject to strong constraints. Other flavor patterns can give larger effects in branching ratios for decays into dimuons, dineutrinos and LFV ones, 
see Table \ref{tab:LQ_branching_fractions}. 
The CP-asymmetry in the $D^+ \to \pi^+ \mu^+ \mu^-$ rate provides an opportunity to probe new physics even for rather suppressed couplings in case of
leptoquarks coupling to $SU(2)$-doublet quarks, see the lower two plots in Fig.~\ref{fig:ACP}.
 Such asymmetries may show up, for instance, with leptoquarks $S_3(3,3,-1/3)$ with electron couplings sufficiently suppressed,  a model that
  can also accommodate recent LNU hints in rare $B \to K l^+ l^-$ decays \cite{Hiller:2014yaa,Gripaios:2014tna}.
 
The benchmark patterns studied in this work  do not exhaust the flavor model space.
We emphasize the importance of searches for FCNCs into dineutrinos and LFV, including $D^0 \to \tau^\pm e^\mp$ decays.
 Further experimental and theoretical study is needed to progress with the quest for BSM and flavor physics.

\bigskip
Note added:
Soon after we published this paper on the arXiv a related analysis \cite{Fajfer:2015mia} appeared.\\
Note also that the recent LHCb bound ${\cal{B}}(D^0 \to e^\pm \mu^\mp )< 1.3 \cdot 10^{-8}$ at 90 \% CL \cite{Aaij:2015qmj} that appeared while this paper has been under review starts to constrain certain leptoquark flavor scenarios, see Table \ref{tab:LQ_branching_fractions}.
Furthermore, the measurement of ${\cal{B}}(D^+ \to \omega \pi^+)$ reported in a recent preprint by BES III \cite{Ablikim:2015wel} yields  $a_\omega=  0.032^{+0.006}_{-0.007} \, \mbox{GeV}^2$, somewhat lower than isospin prediction reported around Eq.~(\ref{eq:C9R_parameters}).

\section*{Acknowledgements}
We thank Christoph Bobeth, Zoltan Ligeti, Bastian M\"uller and Dirk Seidel for useful comments on the manuscript.
GH is grateful to the Kavli Institute for Theoretical Physics  at Santa Barbara where parts of this project have been done for kind hospitality.
This project is supported in part
by the DFG Research Unit FOR 1873 ``Quark Flavour
Physics and Effective Field Theories’' and the National Science Foundation under Grant No. NSF PHY11-25915.

\appendix

\section{Parameters}\label{app:parameters}
$\overline {MS}$ masses are taken from \cite{Agashe:2014kda}
\begin{align}
 &m_t(m_t)=160_{-4}^{+5}\;\text{GeV}\,,\quad&&m_b(m_b)=4.18\pm0.03\;\text{GeV}\,,\\
 &m_c(m_c)=1.275\pm0.025\;\text{GeV}\,,\quad&&m_s(2\;\text{GeV})=0.095\pm0.005\;\text{GeV}\,.
\end{align}
The NNLO running, decoupling at flavor thresholds and quark pole mass are taken from  \cite{Chetyrkin:2000yt}.
The CKM matrix is given by the UTfit collaboration \cite{UTfit}.
The inclusive semileptonic branching fractions are given by the PDG \cite{Agashe:2014kda},
where we use $\mathcal B(D\to Xl^+\nu_l)=\mathcal B(D\to Xl^+)$, consistent with  \cite{Asner:2009pu}, and employ $\mathcal B(D_s^+\to X\mu^+\nu_\mu)\simeq\mathcal B(D_s^+\to Xe^+\nu_e)$.
The particle masses, widths and branching fractions are given by the PDG \cite{Agashe:2014kda}.
The decay constants are given by the FLAG \cite{Aoki:2013ldr}
$f_D=0.2092\pm0.0033\,\text{GeV}\,$, 
 $f_{D_s^+}=0.2486\pm0.0027\,\text{GeV}\,$.
 The bag parameter is \cite{Carrasco:2015pra}
$ B_{D^0}(\mu=3\,\text{GeV})=0.757\pm0.028 $.
We update the nuclear weak charge of cesium \cite{Porsev:2010de,Dzuba:2012kx} using  \cite{Agashe:2014kda} $\Delta Q_w(\text{Cs})=0.69\pm0.44$,
where $\Delta Q_w=Q_w^\text{exp}-Q_w^\text{SM}$.

The leptonic pion decay ratio $R_{e/\mu}^\text{exp}=\Gamma(\pi^+\to(e^+\nu_e+e^+\nu_e\gamma))/\Gamma(\pi^+\to(\mu^+\nu_\mu+\mu^+\nu_\mu\gamma))=(1.230\pm0.004)\cdot10^{-4}$ \cite{Agashe:2014kda}, $R_{e/\mu}^\text{SM}=(1.2352\pm0.0001)\cdot10^{-4}$ \cite{Cirigliano:2007xi} and, thus we find
$ \Delta R_{e/\mu}=(-5.0\pm4.0)\cdot10^{-7}$.
The anomalous magnetic moment of the electron is \cite{Aoyama:2014sxa}
$ \Delta a(e)=(-0.91\pm0.82)\cdot10^{-12}$.
Moreover \cite{Agashe:2014kda}
\begin{align}
 &\Delta a(\mu)=(288\pm63\pm49)\cdot10^{-11}\,,\\
 &|m_{D_1^0}-m_{D_2^0}|=(0.95_{-0.44}^{+0.41})\cdot10^{10}\;\text s^{-1}\,,\\
 &\mathcal B(D^+\to\mu^+\nu_\mu)=(3.82\pm0.33)\cdot10^{-4}\,,\\
 &\mathcal B(D_s^+\to\mu^+\nu_\mu)=(5.56\pm0.25)\cdot10^{-3}
\end{align}
and at CL=90\% \cite{Agashe:2014kda}
\begin{align}
 &d(n)<0.29\cdot10^{-25}\,e\;\text{cm}\,,\\
 &d(e)<10.5\cdot10^{-28}\,e\;\text{cm}\,,\\
 &\mathcal B(\pi^+\to\mu^+\nu_e)<8.0\cdot10^{-3}\,,\\
 &\mathcal B(\mu^-\to e^-\gamma)<5.7\cdot10^{-13}\,,\\
 &\mathcal B(\mu^-\to e^-e^+e^-)<1.0\cdot10^{-12}\,,\\
 &\Gamma(\mu^-\text{Ti}\to e^-\text{Ti})/\Gamma_\text{capture}(\mu^-\text{Ti})<4.3\cdot10^{-12}\,,\\
 &\Gamma(\mu^-\text{Au}\to e^-\text{Au})/\Gamma_\text{capture}(\mu^-\text{Au})<7\cdot10^{-13}\,,
\end{align}
where 
 $\Gamma_\text{capture}(\mu^-\text{Ti})=2.59\cdot10^6\,s^{-1}$ and 
 $\Gamma_\text{capture}(\mu^-\text{Au})=13.07\cdot10^6\,s^{-1}$ \cite{Suzuki:1987jf}.

\section{Effective Wilson Coefficients}\label{app:effective_wilson_coefficients}
In this appendix we give auxiliary  functions and coefficients of the effective Wilson coefficients defined in Sec.~\ref{sec:wilson_coefficients}.
We find
\begin{align}\label{eq:one_loop_qbarq}
 L(m^2,q^2)=\frac53+\ln\frac{\mu_c^2}{m^2}+x-\frac12(2+x)|1-x|^{1/2}
 \begin{cases}
  \ln\frac{1+\sqrt{1-x}}{1-\sqrt{1-x}}-i\pi\quad x\equiv\frac{(2m)^2}{q^2}<1\\
  2\tan^{-1}\left[\frac1{\sqrt{x-1}}\right]\quad x\equiv\frac{(2m)^2}{q^2}>1
 \end{cases}
\end{align}
and in the limit $m^2=0$
\begin{align}
 L(0,q^2)=\frac53+\ln\frac{\mu_c^2}{q^2}+i\pi\,.
\end{align}
We take from \cite{Greub:1996wn}
\begin{align}
 f(\rho)&=-\frac1{243}\big((3672-288\pi^2-1296\zeta_3+(1944-324\pi^2)\ln\rho+108\ln^2\rho+36\ln^3\rho)\rho\nonumber\\
 &+576\pi^2\rho^\frac32+(324-576\pi^2+(1728-216\pi^2)\ln\rho+324\ln^2\rho+36\ln^3\rho)\rho^2\nonumber\\
 &+(1296-12\pi^2+1776\ln\rho-2052\ln^2\rho)\rho^3\big)\nonumber\\
 &-\frac{4\pi i}{81}\big((144-6\pi^2+18\ln\rho+18\ln^2\rho)\rho+(-54-6\pi^2+108\ln\rho+18\ln^2\rho)\rho^2\nonumber\\
 &+(116-96\ln\rho)\rho^3\big)\nonumber\\
 &-\frac{92}{81}\ln\frac{\mu_c^2}{m_c^2}+\frac{983}{243}+\frac{52}{81}\pi i+\mathcal O\left((\rho\ln\rho)^4\right)\,,
\end{align}
where we find the constant terms from \cite{Greub:1996tg}. {}From \cite{Gambino:2003zm} we obtain
\begin{align}
 &y^{(7)}=\left\{0,0,\frac23,\frac89,\frac{40}3,\frac{160}9\right\} \, , \quad \quad 
 y^{(8)}=\left\{0,0,1,-\frac16,20,-\frac{10}3\right\}
\end{align}
and \cite{Greub:2008cy}
\begin{align}
 F_8^{(7)}(\rho)&=\frac{8\pi^2}{27}\frac{(2+\rho)}{(1-\rho)^4}-\frac89\frac{\left(11-16\rho+8\rho^2\right)}{(1-\rho)^2}-\frac{16}9\frac{\sqrt\rho\sqrt{4-\rho}}{(1-\rho)^3}\left(9-5\rho+2\rho^2\right)\arcsin\frac{\sqrt\rho}2\nonumber\\
 &-\frac{32}3\frac{(2+\rho)}{(1-\rho)^4}\arcsin^2\frac{\sqrt\rho}2-\frac{16}9\frac\rho{(1-\rho)}\ln\rho-\frac{32}9\ln\frac{\mu_c^2}{m_c^2}-\frac{16}9\pi i\,,\\
 F_8^{(9)}(\rho)&=-\frac{16\pi^2}{27}\frac{(4-\rho)}{(1-\rho)^4}+\frac{16}9\frac{(5-2\rho)}{(1-\rho)^2}+\frac{32}9\frac{\sqrt{4-\rho}}{\sqrt\rho(1-\rho)^3}\left(4+3\rho-\rho^2\right)\arcsin\frac{\sqrt\rho}2\nonumber\\
 &+\frac{64}3\frac{(4-\rho)}{(1-\rho)^4}\arcsin^2\frac{\sqrt\rho}2+\frac{32}9\frac1{(1-\rho)}\ln\rho\,.
\end{align}

\section{Form Factors}\label{app:form_factors}

\begin{figure}[!htb]
 \centering
 \includegraphics[width=0.8\textwidth]{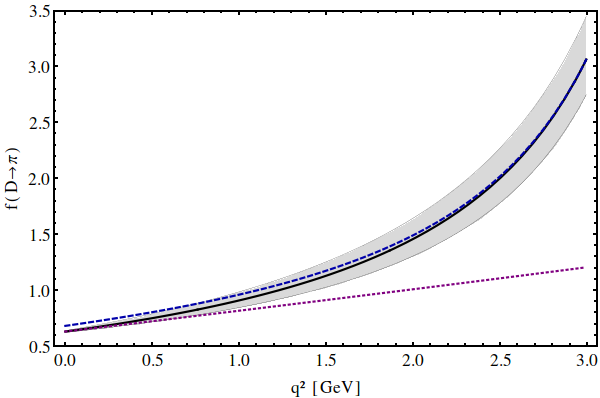}
 \caption{
 The solid black line denotes $f_+$ within its gray uncertainty band, the dashed blue curve denotes $f_T(\mu_c=m_c)$ as derived from Eqs.~(\ref{eq:fT_largerecoil}), (\ref{eq:fT_lowrecoil}) and the dotted purple curve denotes $f_0$ as calculated on the lattice \cite{Koponen:2013ila}.
 Uncertainties for $f_T$ that follow from the parametric ones of $f_+$ are not shown to avoid clutter, but are included in our analysis.}
 \label{fig:form_factors}
\end{figure}

We parametrize the hadronic matrix elements in terms of the form factors $f_i(q^2)$, $i=+,T,0$,
\begin{align}
 &\langle P(p_P)|\bar u\gamma^\mu c|D(p_D)\rangle=f_+(q^2)\left(p^\mu-\frac{m_D^2-m_P^2}{q^2}q^\mu\right)+f_0(q^2)\frac{m_D^2-m_P^2}{q^2}q^\mu\label{eq:me_vector_current}\,,\\
 &\langle P(p_P)|\bar u\sigma^{\mu\nu}(1\pm\gamma_5)c|D(p_D)\rangle=i\frac{f_T(q^2)}{m_D}(p^\mu q^\nu-q^\mu p^\nu\pm i\epsilon^{\mu\nu\rho\sigma}p_\rho q_\sigma)\,,\label{eq:me_tensor_current}
\end{align}
where $q^\mu=(p_D-p_P)^\mu=(p_{l^+}+p_{l^-})^\mu$ and $p^\mu=(p_D+p_P)^\mu$. For $D^0\to\pi^0$ the form factors are scaled $f_i\to f_i/\sqrt2$ by  isospin.
The heavy-to-light form factors are related within the Heavy Quark Effective Theory by means of a heavy quark spin symmetry \cite{Isgur:1990kf,Charles:1998dr}.
At low recoil  \cite{Bobeth:2011nj}
\begin{align}
 &f_T(q^2)=\frac{m_D^2}{q^2}\left(1-\frac{\alpha_s}\pi\frac13\ln\frac{\mu_c^2}{m_c^2}\right)f_+(q^2)+\mathcal O\left(\frac{\Lambda_\text{QCD}}{m_c},\alpha_s^2\right)\label{eq:fT_lowrecoil}\,.
\end{align}
The breaking of the heavy quark spin symmetry at large recoil reads \cite{Beneke:2000wa}
\begin{align}
 &f_T(q^2)=\left(1+\frac{\alpha_s}\pi\left(-\frac23\frac{2E}{m_D-2E}\ln\frac{2E}{m_D}-\frac13\ln\frac{\mu_c^2}{m_c^2}\right)\right)f_+(q^2)\label{eq:fT_largerecoil}\,,
\end{align}
where $E=(m_D^2-m_P^2-q^2)/(2m_D)$.
In our analysis we interpolate between (\ref{eq:fT_largerecoil}) and (\ref{eq:fT_lowrecoil}) and take $f_0$ from a lattice calculation \cite{Koponen:2013ila}.
For  the residual form factor we use  the $z$-expansion \cite{Amhis:2014hma}
\begin{align}
 f_+(q^2)=\frac1{\phi(q^2,t_0)}\sum_{i=0}^\infty a_i(t_0)\left(z(q^2,t_0)\right)^i \, ,
\end{align}
with
\begin{align}
 &z(q^2,t_0)=\frac{\sqrt{t_+-q^2}-\sqrt{t_+-t_0}}{\sqrt{t_+-q^2}+\sqrt{t_+-t_0}}\,,  \quad 
 t_\pm=(m_D\pm m_P)^2\,, \quad 
 t_0=t_+\left(1-\sqrt{1-\frac{t_-}{t_+}}\right)\,, \\
 &\phi(q^2,t_0)=\sqrt{\frac{\pi m_c^2}3}\left(\sqrt{t_+-q^2}+\sqrt{t_+-t_0}\right)\frac{t_+-q^2}{(t_+-t_0)^{1/4}}\frac{\left(\sqrt{t_+-q^2}+\sqrt{t_+-t_-}\right)^{3/2}}{\left(\sqrt{t_+-q^2}+\sqrt{t_+}\right)^5}\,.
\end{align}
Assuming isospin symmetry, we employ the parameters to second order as given by HFAG \cite{Amhis:2014hma} 
\begin{align}
 f_+(0)\left|V_{cd}\right|=0.1425\pm0.0019\,,\quad r_1=-1.94\pm0.19\,,\quad r_2=-0.62\pm1.19 \, ,
\end{align}
where $r_i\equiv a_i/a_0$ and $m_i=(m_{i^+}+m_{i^0})/2$.
Lattice computations for $f_+(q^2)$ \cite{Koponen:2013ila} are  consistent with \cite{Amhis:2014hma}, and find insensitivity of $f_+$ to the spectator quark.
We therefore use identical numerics for $D \to \pi$ and $D_s \to K$ form factors.
The  form factors  as used in our analysis are shown in Fig.~\ref{fig:form_factors}. We do not take into account uncertainties in $f_0$, which are $\lesssim10\%$ \cite{Koponen:2013ila} as this enters BSM predictions only, and
because they are negligible in view of other uncertainties.

\section{Exclusive Charm Decay Observables}\label{app:observables}

Here we give the observables for exclusive charm decays used in our analysis.
The form factors $f_i$ are defined in App.~\ref{app:form_factors}. We neglect non-factorizable terms.
The $D\to Pll$ distributions are in agreement with \cite{Bobeth:2007dw}. The dilepton spectrum reads
\begin{align}\label{eq:dGammaq2_eob}
 \frac{\mathrm d\Gamma}{\mathrm dq^2}&=\frac{G_F^2\alpha_e^2}{1024\pi^5m_D^3}\bigg(\frac23\left(\left(|C_9|^2+|C_{10}|^2\right)f_+^2+4|C_7|^2f_T^2\frac{m_c^2}{m_D^2}+4\mathrm{Re}\left[C_7C_9^*\right]f_Tf_+\frac{m_c}{m_D}\right)\nonumber\\
 &\times\lambda(m_D^2,m_P^2,q^2)\left(1+\frac{2m_l^2}{q^2}\right)+|C_{10}|^2\left(-f_+^2\lambda(m_D^2,m_P^2,q^2)+f_0^2\left(m_D^2-m_P^2\right)^2\right)\frac{4m_l^2}{q^2}\nonumber\\
 &+\left(|C_S|^2\left(q^2-4m_l^2\right)+|C_P|^2q^2\right)f_0^2\frac{\left(m_D^2-m_P^2\right)^2}{m_c^2}\nonumber\\
 &+\frac43\left(|C_T|^2+|C_{T5}|^2\right)f_T^2\frac{q^2\lambda(m_D^2,m_P^2,q^2)}{m_D^2}\left(1-\frac{4m_l^2}{q^2}\right)\nonumber\\
 &+8\mathrm{Re}\left[\left(C_9f_++2C_7f_T\frac{m_c}{m_D}\right)C_T^*\right]f_T\frac{\lambda(m_D^2,m_P^2,q^2)}{m_D}m_l\nonumber\\
 &+4\mathrm{Re}\left[C_{10}C_P^*\right]f_0^2\frac{\left(m_D^2-m_P^2\right)^2}{m_c}m_l\nonumber\\
 &+16|C_T|^2f_T^2\frac{\lambda(m_D^2,m_P^2,q^2)}{m_D^2}m_l^2\bigg)\sqrt{\lambda(m_D^2,m_P^2,q^2)\left(1-\frac{4m_l^2}{q^2}\right)} \, ,
\end{align}
where $ \lambda(a,b,c)=a^2+b^2+c^2-2ab-2ac-2bc$.
The differential lepton forward-backward asymmetry defined as the asymmetry between forward minus backward flying $l^-$  in the dilepton center of mass frame relative to the recoiling $P$ reads
\begin{align}\label{eq:AFB}
 A_{\rm FB}(q^2)& 
  =N\frac{G_F^2\alpha_e^2}{512\pi^5m_D^3}\bigg(\mathrm{Re}\left[\left(C_SC_T^*+C_PC_{T5}^*\right)\right]f_T\frac{m_D^2-m_P^2}{m_c\,m_D}q^2\nonumber\\
 &+\mathrm{Re}\left[\left(C_9f_++2C_7f_T\frac{m_c}{m_D}\right)C_S^*\right]\frac{m_D^2-m_P^2}{m_c}m_l\nonumber\\
 &+2\mathrm{Re}\left[C_{10}C_{T5}^*\right]f_T\frac{m_D^2-m_P^2}{m_D}m_l\bigg)f_0\lambda(m_D^2,m_P^2,q^2)\left(1-\frac{4m_l^2}{q^2}\right)\,.
\end{align}
For  vanishing lepton masses  the flat term  \cite{Bobeth:2007dw} reads
\begin{align}\label{eq:FH}
 F_H(q^2)&=N\frac{G_F^2\alpha_e^2}{2048\pi^5m_D^3}\bigg(\left(|C_S|^2+|C_P|^2\right)f_0^2\frac{\left(m_D^2-m_P^2\right)^2}{m_c^2}\nonumber\\
 &+4\left(|C_T|^2+|C_{T5}|^2\right)f_T^2\frac{\lambda(m_D^2,m_P^2,q^2)}{m_D^2}\bigg)q^2\sqrt{\lambda(m_D^2,m_P^2,q^2)}+\mathcal O(m_l)\,,
\end{align}
where $N^{-1}=\int_{q^2_\text{min}}^{q^2_\text{max}}\mathrm dq^2\,\mathrm d\Gamma/\mathrm dq^2$.
For the LFV $D\to Pe\mu$ decay distributions we obtain, for $m_e=0$,
\begin{align}\label{eq:dGammaq2_DPemu}
 \frac{\mathrm d\Gamma (D^+\to P^+e^\pm\mu^\mp)}{\mathrm dq^2}&=\frac{G_F^2\alpha_e^2}{1024\pi^5m_D^3}  \sqrt{\lambda(m_D^2,m_P^2,q^2)}   \bigg(\frac23\left(|K_9|^2+|K_{10}|^2\right)f_+^2\lambda(m_D^2,m_P^2,q^2)\nonumber\\
 &+\left(|K_S|^2+|K_P|^2\right)f_0^2\frac{\left(m_D^2-m_P^2\right)^2}{m_c^2}q^2\nonumber\\
 &+\frac43\left(|K_T|^2+|K_{T5}|^2\right)f_T^2\frac{q^2\lambda(m_D^2,m_P^2,q^2)}{m_D^2}\nonumber\\
 &+2\mathrm{Re}\left[\pm K_9K_S^*+K_{10}K_P^*\right]f_0^2\frac{\left(m_D^2-m_P^2\right)^2}{m_c}m_\mu\nonumber\\
 &+4\mathrm{Re}\left[K_9K_T^*\pm K_{10}K_{T5}^*\right]f_Tf_+\frac{\lambda(m_D^2,m_P^2,q^2)}{m_D}m_\mu\bigg) 
+\mathcal O\left(m_\mu^2\right)\,,
\end{align}
where $K_i=K_i^{(\mu)}$ and the plus signs for $D^+\to P^+e^+\mu^-$, and $K_i=K_i^{(e)}$ and the minus signs for $D^+\to P^+e^+\mu^-$.
Compared to Eq.~(\ref{eq:dGammaq2_eob}), additional vector-scalar and axialvector-axialtensor are present in Eq.~(\ref{eq:dGammaq2_DPemu}).

All Wilson coefficients except for those of the tensors in Eqs.~(\ref{eq:dGammaq2_eob})-(\ref{eq:dGammaq2_DPemu}) are tacitly understood as $C_i \to C_i +C_i^\prime$ and
$K_i \to K_i +K_i^\prime$, that is, primed Wilson coefficients are added.

The $D^0\to l^+l^-$ branching fraction can be inferred from \cite{Bobeth:2007dw}
\begin{align}\label{eq:B_D0ll}
 \mathcal B(D^0\to l^+l^-)&=\frac{G_F^2\alpha_e^2m_{D^0}^5f_{D^0}^2}{64\pi^3\Gamma_{D^0}}\sqrt{1-\frac{4m_l^2}{m_{D^0}^2}}\bigg(\left(1-\frac{4m_l^2}{m_{D^0}^2}\right)\left|\frac{C_S-C_S'}{m_c}\right|^2\nonumber\\
 &+\left|\frac{C_P-C_P'}{m_c}+\frac{2m_l}{m_{D^0}^2}(C_{10}-C_{10}')\right|^2\bigg) \, .
\end{align}
The LFV ones read, for $m_e=0$,  
\begin{align} \nonumber
 \mathcal B(D^0\to e^- \mu^+)&=\frac{G_F^2\alpha_e^2m_{D^0}^5f_{D^0}^2}{64\pi^3\Gamma_{D^0} }\left(1-\frac{m_\mu^2}{m_{D^0}^2}\right)^2 \bigg( \left|\frac{K_S^{(e)}-K_S^{(e) \prime}}{m_c} - \frac{ m_\mu }{m_{D^0}^2} \left(K_9^{(e)}-K_9^{(e) \prime}\right) \right|^2\nonumber\\   \nonumber
 &+\left|\frac{K_P^{(e)}-K_P^{(e) \prime}}{m_c}+\frac{m_\mu}{m_{D^0}^2}\left(K_{10}^{(e)}-K_{10}^{(e) \prime}\right)\right|^2\bigg) \, ,  \\
 \mathcal B(D^0\to \mu^- e^+)&=\frac{G_F^2\alpha_e^2m_{D^0}^5f_{D^0}^2}{64\pi^3\Gamma_{D^0}}  \left(1-\frac{m_\mu^2}{m_{D^0}^2}\right)^2  \bigg(\left|\frac{K_S^{(\mu)}-K_S^{(\mu) \prime}}{m_c} + \frac{ m_\mu }{m_{D^0}^2} \left(K_9^{(\mu)}-K_9^{(\mu) \prime}\right) \right|^2\nonumber\\
 &+\left|\frac{K_P^{(\mu)}-K_P^{(\mu) \prime}}{m_c}+\frac{m_\mu}{m_{D^0}^2}\left(K_{10}^{(\mu)}-K_{10}^{(\mu) \prime}\right)\right|^2\bigg) \, .       \label{eq:B_D0mue} 
\end{align}

\section{Non-resonant SM \texorpdfstring{$c\to ull$}{ctoull} Branching Fractions}\label{app:branching_fractions}

In this appendix we provide our predictions for the non-resonant SM branching fractions of the inclusive $c\to ull$ decays and exclusive  $D\to Pll$ modes.
In our analysis uncertainties due to power corrections are not included. Electroweak corrections, which are subleading relative to QCD-ones, are neglected.
For the $D\to Pll$ modes we integrate the branching fractions over different dilepton masses, $\sqrt{q^2}\ge2m_l$ (Table~\ref{tab:DPll_2ml}), $0.250\,\text{GeV}\le\sqrt{q^2}\le0.525\,\text{GeV}$ (Table~\ref{tab:DPll_0250_0525}) and $\sqrt{q^2}\ge1.25\,\text{GeV}$ (Table~\ref{tab:DPll_125}).
\begin{table}[!htb]
 \centering
 \begin{tabular}{c|c|c}
  mode                      &  branching fraction                                                                                 &  90\% CL limit \cite{Agashe:2014kda}  \\
  \noalign{\hrule height 1pt}
  $D^+\to\pi^+e^+e^-$       &  $4.6\cdot10^{-12}\,(\pm1,_{-1}^{+2},_{-13}^{+14},\pm1,_{-1}^{+5},_{-3}^{+210},_{-10}^{+13})$       &  $1.1\cdot10^{-6}$  \\
  \hline
  $D^+\to\pi^+\mu^+\mu^-$   &  $3.7\cdot10^{-12}\,(\pm1,\pm3,_{-15}^{+16},\pm1,_{-1}^{+3},_{-1}^{+158},_{-12}^{+16})$             &  $7.3\cdot10^{-8}$  \\
  \hline
  $D^0\to\pi^0e^+e^-$       &  $9.1\cdot10^{-13}\,(\pm1,\pm1,_{-13}^{+14},\pm1,_{-1}^{+5},_{-2}^{+211},_{-10}^{+13})$             &  $4.5\cdot10^{-5}$  \\
  \hline
  $D^0\to\pi^0\mu^+\mu^-$   &  $7.3\cdot10^{-13}\,(\pm1,\pm3,_{-15}^{+16},\pm1,_{-1}^{+3},_{-1}^{+159},_{-12}^{+16})$             &  $1.8\cdot10^{-4}$  \\
  \hline
  $D_s^+\to K^+e^+e^-$      &  $1.7\cdot10^{-12}\,(\pm2,_{-3}^{+4},_{-12}^{+13},_{-1}^{+7},_{-1}^{+8},_{-7}^{+228},_{-8}^{+10})$  &  $3.7\cdot10^{-6}$  \\
  \hline
  $D_s^+\to K^+\mu^+\mu^-$  &  $1.2\cdot10^{-12}\,(\pm2,_{-1}^{+2},_{-15}^{+16},\pm2,_{-1}^{+4},_{-1}^{+167},_{-10}^{+13})$       &  $2.1\cdot10^{-5}$  \\
 \end{tabular}
 \caption{Non-resonant SM branching fractions for $\sqrt{q^2}\ge2m_l$ of $D\to Pll$ decays normalized to the width.
 Non-negligible uncertainties are labeled by (normalization, $m_c$, $m_s$, $\mu_W$, $\mu_b$, $\mu_c$, $f_+$) given in percentage, where $m_{W,b}/2\le\mu_{W,b}\le2m_{W,b}$ and $m_c/\sqrt2\le\mu_c\le\sqrt2m_c$. }
 \label{tab:DPll_2ml}
\end{table}
\begin{table}[!htb]
 \centering
 \begin{tabular}{c|c|c}
  mode                      &  branching fraction                                                                              &  90\% CL limit \cite{Aaij:2013sua}  \\
  \noalign{\hrule height 1pt}
  $D^+\to\pi^+e^+e^-$       &  $8.1\cdot10^{-13}\,(\pm1,_{-4}^{+5},_{-22}^{+23},_{-12}^{+11},_{-1}^{+10},_{-24}^{+247},\pm5)$  &  --  \\
  \hline
  $D^+\to\pi^+\mu^+\mu^-$   &  $7.4\cdot10^{-13}\,(\pm1,\pm4,_{-21}^{+23},_{-11}^{+10},_{-1}^{+10},_{-23}^{+238},_{-5}^{+6})$  &  $2.0\cdot10^{-8}$  \\
  \hline
  $D^0\to\pi^0e^+e^-$       &  $1.6\cdot10^{-13}\,(\pm1,_{-4}^{+5},_{-22}^{+23},_{-12}^{+11},_{-1}^{+10},_{-24}^{+247},\pm5)$  &  --  \\
  \hline
  $D^0\to\pi^0\mu^+\mu^-$   &  $1.5\cdot10^{-13}\,(\pm1,\pm4,_{-21}^{+23},_{-11}^{+10},_{-1}^{+10},_{-22}^{+238},_{-5}^{+6})$  &  --  \\
  \hline
  $D_s^+\to K^+e^+e^-$      &  $3.6\cdot10^{-13}\,(\pm2,\pm5,_{-22}^{+23},_{-13}^{+12},_{-1}^{+11},_{-21}^{+248},\pm5)$        &  --  \\
  \hline
  $D_s^+\to K^+\mu^+\mu^-$  &  $3.3\cdot10^{-13}\,(\pm2,\pm5,_{-22}^{+23},_{-13}^{+12},_{-1}^{+11},_{-20}^{+239},_{-5}^{+6})$  &  --  \\
 \end{tabular}
 \caption{ As in Table \ref{tab:DPll_2ml} but for the low $q^2$-region,  $0.250\,\text{GeV}\le\sqrt{q^2}\le0.525\,\text{GeV}$.}
 \label{tab:DPll_0250_0525}
\end{table}
\begin{table}[!htb]
 \centering
 \begin{tabular}{c|c|c}
  mode                  &  branching fraction                                                                       &  90\% CL limit \cite{Aaij:2013sua}  \\
  \noalign{\hrule height 1pt}
  $D^+\to\pi^+l^+l^-$   &  $7.4\cdot10^{-13}\,(\pm1,\pm6,_{-14}^{+15},\pm6,_{-1}^{+0},_{-45}^{+136},_{-20}^{+27})$  &  $2.6\cdot10^{-8}$ ($l=\mu$)  \\
  \hline
  $D^0\to\pi^0l^+l^-$   &  $1.4\cdot10^{-13}\,(\pm1,\pm6,_{-14}^{+15},\pm6,_{-1}^{+0},_{-45}^{+136},_{-20}^{+27})$  &  --  \\
  \hline
  $D_s^+\to K^+l^+l^-$  &  $7.9\cdot10^{-14}\,(\pm2,\pm6,_{-14}^{+15},\pm6,_{-1}^{+0},_{-45}^{+133},_{-19}^{+26})$  &  --  \\
 \end{tabular}
 \caption{As in Table \ref{tab:DPll_2ml} but for the high $q^2$-region,   $\sqrt{q^2}\ge1.25\,\text{GeV}$, and $l=e,\mu$.}
 \label{tab:DPll_125}
\end{table}

Next, we obtain inclusive $c\to ull$  branching fractions.
To leading order in the heavy quark expansion and neglecting lepton masses the $q^2$-distribution reads \cite{Fukae:1998qy}
\begin{align}\label{eq:inclusive_distribution}
 \frac{\mathrm d\Gamma (c\to ull)}{\mathrm dq^2}&=\frac{G_F^2\alpha_e^2m_c^3}{768\pi^5}\left(1-\frac{q^2}{m_c^2}\right)^2\bigg[\left(1+2\frac{q^2}{m_c^2}\right)\left(|C_9|^2+|C_{10}|^2\right)+4\left(2\frac{m_c^2}{q^2}+1\right)\left|C_7\right|^2\nonumber\\
 &+12\mathrm{Re}\left[C_7C_9^*\right]\bigg]\,,
\end{align}
where $q^2=(p_c-p_u)^2=(p_{l^+}+p_{l^-})^2$ and $(2m_l)^2\le q^2\le m_c^2$.

The matrix elements at NLO QCD are obtained as $C_i\to C_i(1+\alpha_s/\pi\,\sigma_i(q^2/m_c^2))$ \cite{Ghinculov:2003qd} (and references therein)
\begin{align}
 &\sigma_7(\rho)=-\frac43\mathrm{Li_2}[\rho]-\frac23\ln\rho\ln[1-\rho]-\frac29\pi^2-\ln[1-\rho]-\frac29(1-\rho)\ln[1-\rho]+\frac16-\frac43\ln\frac{\mu_c^2}{m_c^2}\,,\\
 &\sigma_9(\rho)=-\frac43\mathrm{Li_2}[\rho]-\frac23\ln\rho\ln[1-\rho]-\frac29\pi^2-\ln[1-\rho]-\frac29(1-\rho)\ln[1-\rho]+\frac32\,,
\end{align}
where $\sigma_{10}=\sigma_9$ and via $\mathrm{Re}[C_iC_j^*]\to\mathrm{Re}[C_iC_j^*](1+\alpha_s/\pi\,\tau_{ij}^{(1)}(q^2/m_c^2))$
\begin{align}
 \tau_{77}^{(1)}(\rho)&=-\frac2{9(2+\rho)}\left(2(1-\rho)^2\ln[1-\rho]+\frac{6\rho(2-2\rho-\rho^2)}{(1-\rho)^2}\ln\rho+\frac{11-7\rho-10\rho^2}{1-\rho}\right)\,,\\
 \tau_{99}^{(1)}(\rho)&=-\frac4{9(1+2\rho)}\left(2(1-\rho)^2\ln[1-\rho]+\frac{3\rho(1+\rho)(1-2\rho)}{(1-\rho)^2}\ln\rho+\frac{3(1-3\rho^2}{1-\rho}\right)\,,\\
 \tau_{79}^{(1)}(\rho)&=-\frac{4(1-\rho)^2}{9\rho}\ln[1-\rho]-\frac{4\rho(3-2\rho)}{9(1-\rho)^2}\ln\rho-\frac{2(5-3\rho)}{9(1-\rho)}\,,\\
 \tau_{710}^{(1)}(\rho)&=-\frac52+\frac1{3(1-3\rho)}-\frac{\rho(6-7\rho)\ln\rho}{3(1-\rho)^2}-\frac{(3-7\rho+4\rho^2)\ln[1-\rho]}{9\rho}\nonumber\\
 &+\frac1{18(1-\rho)^2}\big[24\left(1+13\rho-4\rho^2\right)\mathrm{Li_2}[\sqrt\rho]+12\left(1-17\rho+6\rho^2\right)\mathrm{Li_2}[\rho]+6\rho(6-7\rho)\ln\rho\nonumber\\
 &+24(1-\rho)^2\ln\rho\ln[1-\rho]+12\left(-13+16\rho-3\rho^2\right)(\ln[1-\sqrt\rho]-\ln[1-\rho])\nonumber\\
 &+39-2\pi^2+252\rho-26\pi^2\rho+21\rho^2+8\pi^2\rho^2-180\sqrt\rho-132\rho\sqrt\rho\big]\,,\\
 \tau_{910}^{(1)}(\rho)&=-\frac52+\frac1{3(1-\rho)}-\frac{\rho(6-7\rho)\ln\rho}{3(1-\rho)^2}-\frac{2(3-5\rho+2\rho^2)\ln[1-\rho]}{9\rho}\nonumber\\
 &-\frac1{18(1-\rho)^2}\big[48\rho(-5+2\rho)\mathrm{Li_2}[\sqrt\rho]+24\left(-1+7\rho-3\rho^2\right)\mathrm{Li_2}[\rho]+6\rho(-6+7\rho)\ln\rho\nonumber\\
 &-24(1-\rho)^2\ln\rho\ln[1-\rho]+24\left(5-7\rho+2\rho^2\right)(\ln[1-\sqrt\rho]-\ln[1-\rho])\nonumber\\
 &-21-156\rho+20\pi^2\rho+9\rho^2-8\pi^2\rho^2+120\sqrt\rho+48\rho\sqrt\rho\big]\,,
\end{align}
where $\tau_{10\,10}=\tau_{99}$.
We obtain the NNLO term $\delta^{(2)}|C_9|^2=|C_9|^2(\alpha_s/\pi)^2\,\tau_{99}^{(2)}(q^2/m_c^2)$ as \cite{Blokland:2005vq}
\begin{align}
 \tau_{99}^{(2)}(\rho)&=\frac1{\left(1-\rho\right)^2\left(1+2\rho\right)}\bigg[2\left(2.854-0.665\rho-0.109\rho^2-8.572\rho^3+5.561\rho^4+0.931\rho^5\right)\nonumber\\
 &+\frac23\left(-0.063615+0.098146\rho+0.144642\rho^2-0.307331\rho^3+0.107417\rho^4+0.020707\rho^5\right)\nonumber\\
 &+\frac{16}9\left(3.575-2.867\rho+2.241\rho^2-12.027\rho^3+11.564\rho^4-2.489\rho^5\right)\nonumber\\
 &+4\left(-8.151+2.990\rho-3.537\rho^2+36.561\rho^3-42.275\rho^4+23.899\rho^5-9.494\rho^6\right)\bigg] \, .
\end{align}
We normalize  to the $c\to(d,s)l\nu$ width and the experimental branching fraction
\begin{align}\label{eq:cull_branching_fraction}
 \frac{\mathrm d\mathcal B_{D\to X_ull}}{\mathrm dq^2}=\frac{\mathcal B(D\to X_{(d,s)}l\nu)}{\Gamma_{c\to(d,s)l\nu}}\frac{\mathrm d\Gamma_{c\to ull}}{\mathrm dq^2}
\end{align}
with \cite{Pak:2008cp}
\begin{align}
 \Gamma_{c\to(d,s)l\nu}&=\frac{G_F^2m_c^3}{192\pi^3}\sum_{q\in\{d,s\}}|V_{cq}|^2\left(X_0(m_q/m_c)+\frac{\alpha_s}\pi X_1(m_q/m_c)+\left(\frac{\alpha_s}\pi\right)^2X_2(m_q/m_c)\right)\,,
\end{align}
where the functions $X_i$ are given in \cite{Pak:2008cp}.
Power corrections can be inferred from \cite{Ali:1996bm,Buchalla:1998mt}. They are, however,  not included in our numerical analysis, as the
OPE breaks down for large $q^2$ when the inclusive decay ceases to be inclusive but rather  degenerates into few exclusive modes.
 Yet, the power corrections in the region where the OPE works are a small effect  on the uncertainty budget at low $q^2$. 
A comprehensive treatment of the full $q^2$-region is beyond the scope of this work.
Our resulting  inclusive $c\to ull$ branching fractions are compiled in Table \ref{tab:incl}.
\begin{table}[!htb]
 \centering
 \begin{tabular}{c|c}
  mode                        &  branching fraction  \\
  \noalign{\hrule height 1pt}
  $D^+\to X_u^+e^+e^-$        &  $9.4\cdot10^{-10}\,(\pm2,_{-6}^{+7},_{-14}^{+16},\pm7,\pm1,_{-43}^{+109})$  \\
  \hline
  $D^+\to X_u^+\mu^+\mu^-$    &  $2.0\cdot10^{-10}\,(\pm19,_{-5}^{+5},_{-14}^{+15},_{-7}^{+8},\pm1,_{-40}^{+120})$  \\
  \hline
  $D^0\to X_u^0e^+e^-$        &  $3.8\cdot10^{-10}\,(\pm2,_{-6}^{+7},_{-14}^{+16},\pm7,\pm1,_{-43}^{+109})$  \\
  \hline
  $D^0\to X_u^0\mu^+\mu^-$    &  $7.7\cdot10^{-11}\,(\pm9,_{-5}^{+5},_{-14}^{+15},_{-7}^{+8},\pm1,_{-40}^{+120})$  \\
  \hline
  $D_s^+\to X_u^+e^+e^-$      &  $3.8\cdot10^{-10}\,(\pm7,_{-6}^{+7},_{-14}^{+16},\pm7,\pm1,_{-43}^{+109})$  \\
  \hline
  $D_s^+\to X_u^+\mu^+\mu^-$  &  $7.5\cdot10^{-11}\,(\pm7,_{-5}^{+5},_{-14}^{+15},_{-7}^{+8},\pm1,_{-40}^{+120})$  \\
 \end{tabular}
 \caption{Non-resonant SM branching fractions for $\sqrt{q^2}\ge2m_l$ of $D\to X_ull$ decays normalized to $D\to X_{d,s}l\nu$ and vanishing lepton masses  except for the lower cut.
 Non-negligible uncertainties are labeled by (normalization, $m_c$, $m_s$, $\mu_W$, $\mu_b$, $\mu_c$) given in percentage, where $m_{W,b}/2\le\mu_{W,b}\le2m_{W,b}$ and $m_c/\sqrt2\le\mu_c\le\sqrt2m_c$.}
 \label{tab:incl}
\end{table}

\section{Leptoquark Constraints}\label{app:LQ_constraints}

In this appendix we provide constraints on the couplings of the scalar and vector leptoquarks of Table~\ref{tab:cull_leptoquarks}.
Collider experiments find $M\gtrsim1\;\text{TeV}$ \cite{Aad:2015caa,CMS:zva}, thus we use $M=1\,\text{TeV}$ as reference.
We neglect RGE effects from $M$ to $\mu_W$ and further to $\mu_c$  noting that $Q_9$ and $Q_{10}$ do not scale and $C_{S,P}(\mu\sim1\,\text{TeV})\simeq0.5\,C_{S,P}(\mu\sim\mu_c)$ and $C_{T,T5}(\mu\sim1\,\text{TeV})\simeq1.3\,C_{T,T5}(\mu\sim\mu_c)$ at one-loop QCD \cite{Bobeth:2007dw}. Neglecting
such effects is within the accuracy aimed at in this work.
We do not constrain non-gauge vector leptoquarks, which could depend on the cutoff-scale within some model \cite{Davidson:1993qk}.
We first list the constraints on the couplings and the related observables for scalar (Tables~\ref{tab:LQ_S12_constraints} and \ref{tab:LQ_S3_constraints}) and vector (Table~\ref{tab:LQ_V_constraints}) leptoquarks.
\begin{table}
 \centering
 \begin{tabular}{c|c|c}
  couplings/mass                                                                                                     &  constraint                  &  observable  \\
  \noalign{\hrule height 1pt}
  $|\lambda_{S_1L}^{(ue)}|$                                                                                          &  $\lesssim9\cdot10^{-2}$     &  $V_{ud}$  \\
  \hline
  $\lambda_L^{(ue)}\lambda_R^{(ue)}$                                                                                 &  $\sim[-0.08,0.8]$           &  nuclear beta decay  \\
  \hline
  $|\lambda_L^{(ue)}\lambda_R^{(u\mu)}|$                                                                             &  $\lesssim9\cdot10^{-2}$     &  $\pi^+\to\mu^+\nu_e$  \\
  \hline
  $|\lambda_{\{L,R\}}^{(ue)}\lambda_{\{L,R\}}^{(ce)}|$                                                               &  $\lesssim1\cdot10^{-1}$     &  $D^+\to\pi^+e^+e^-$  \\
  \hline
  $|\lambda_{\{L,R\}}^{(ue)}\lambda_{\{L,R\}}^{(c\mu)}|$                                                             &  $\lesssim2\cdot10^{-1}$     &  $D^+\to\pi^+e^+\mu^-$  \\
  \hline
  $|\lambda_{\{L,R\}}^{(u\mu)}\lambda_{\{L,R\}}^{(ce)}|$                                                             &  $\lesssim2\cdot10^{-1}$     &  $D^+\to\pi^+e^-\mu^+$  \\
  \hline
  $|\lambda_{\{L,R\}}^{(u\mu)}\lambda_{\{L,R\}}^{(c\mu)}|$                                                           &  $\lesssim2\cdot10^{-2}$     &  $D^+\to\pi^+\mu^+\mu^-$  \\
  \hline
  $\lambda_L^{(ce)}\lambda_R^{(ce)}$                                                                                 &  $\sim[-0.005,0.05]$         &  $D_s^+\to\mu^+\nu_\mu$\\
  \hline
  $\lambda_{S_1L}^{(ce)}\lambda_{S_1R}^{(ce)}$                                                                       &  $\sim[-0.01,0.00]$          &  $\Delta a_e$  \\
  $\lambda_{S_2L}^{(ce)}\lambda_{S_2R}^{(ce)}$                                                                       &  $\sim[0.00,0.01]$           &  \\
  \hline
  $\lambda_{S_1L}^{(c\mu)}\lambda_{S_1R}^{(c\mu)}$                                                                   &  $\sim[0.1,0.2]$             &  $\Delta a_\mu$  \\
  $\lambda_{S_2L}^{(c\mu)}\lambda_{S_2R}^{(c\mu)}$                                                                   &  $\sim[-0.2,-0.1]$           &  \\
  \hline  
  $(-|\lambda_{S_1L,S_2L}^{(ue)}|^2+|\lambda_{S_1R,S_2R}^{(ue)}|^2+|\lambda_{S_2R}^{(ue)}|^2)^{1/2}$                 &  $\sim[0.2,0.4]$             &  $Q_w\,(\text{Cs})$  \\
  \hline
  $(\lambda_L^{(ue)}\lambda_R^{(ue)}-0.02\,\lambda_L^{(u\mu)}\lambda_R^{(u\mu)}$                                     &                              &  $\Delta R_{e/\mu}$  \\
  $-0.0002\,(|\lambda_{S_1L}^{(ue)}|^2-|\lambda_{S_1L}^{(u\mu)}|^2))$                                         &  $\sim[-0.00001,-0.000003]$  &  \\
  \hline
  $(\lambda_{S_1L,S_2R}^{(ue)}\lambda_{S_1R,S_2L}^{(ce)}-0.02\,\lambda_{S_1L}^{(ue)}\lambda_{S_1L}^{(ce)})$          &  $\sim[-0.009,0.01]$         &  $D^+\to\mu^+\nu_\mu$  \\
  \hline
  $|\lambda_L^{(ue)}\lambda_R^{(ce)}\pm\lambda_R^{(ue)}\lambda_L^{(ce)}|$                                            &  $\lesssim1\cdot10^{-2}$     &  $D^0\to e^+e^-$  \\
  \hline
  $\lambda_{L,R}^{(ue)}\lambda_{L,R}^{(ce)}+\lambda_{L,R}^{(u\mu)}\lambda_{L,R}^{(c\mu)}$                            &  $\sim[0,0.01]$               &  $\Delta m_{D^0}$  \\
  \hline
  $|\lambda_{S_1L,S_1R}^{(qe)}\lambda_{S_1L,S_1R}^{(q\mu)}|$                                                         &  $\lesssim1\cdot10^{-3}$     &  $\mu^-\to e^-\gamma$  \\
  $|\lambda_{S_2L,S_2R}^{(qe)}\lambda_{S_2L,S_2R}^{(q\mu)}|$                                                         &  $\lesssim5\cdot10^{-4}$     &  \\
  $|\lambda_{L,R}^{(ce)}\lambda_{R,L}^{(c\mu)}|$                                                                     &  $\lesssim3\cdot10^{-6}$     &  \\
  \hline
  $|\lambda_{L,R}^{(ue)}\lambda_{L,R}^{(u\mu)}|$                                                                     &  $\lesssim9\cdot10^{-7}$     &  $\mu-e\,(\text{Au})$  \\
  $|\lambda_{S_2R}^{(ue)}\lambda_{S_2R}^{(u\mu)}|$                                                                   &  $\lesssim7\cdot10^{-7}$     &  \\
  $|\lambda_{L,R}^{(ue)}\lambda_{R,L}^{(u\mu)}|$                                                                     &  $\lesssim4\cdot10^{-7}$     &  \\
  $|\lambda_{S_1L,S_1R}^{(ce)}\lambda_{S_1L,S_1R}^{(c\mu)}|$                                                         &  $\lesssim9\cdot10^{-3}$     &  \\
  $|\lambda_{S_2L,S_2R}^{(ce)}\lambda_{S_2L,S_2R}^{(c\mu)}|$                                                         &  $\lesssim3\cdot10^{-3}$     &  \\
  $|\lambda_{L,R}^{(ce)}\lambda_{R,L}^{(c\mu)}|$                                                                     &  $\lesssim1\cdot10^{-5}$     &  \\
  \hline
  $|\lambda_{L,R}^{(ue)}\lambda_{L,R}^{(u\mu)}|$                                                                     &  $\lesssim1\cdot10^{-3}$     &  $\mu^-\to e^-e^+e^-$  \\
  $|\lambda_{S_2R}^{(ue)}\lambda_{S_2R}^{(u\mu)}|$                                                                   &  $\lesssim3\cdot10^{-3}$     &  \\
  $|\lambda_{L,R}^{(ue)}\lambda_{R,L}^{(u\mu)}|$                                                                     &  $\lesssim1\cdot10^{-2}$     &  \\
  $|\lambda_{L,R}^{(ce)}\lambda_{L,R}^{(c\mu)}|$                                                                     &  $\lesssim3\cdot10^{-3}$     &  \\
  $|\lambda_{S_2R}^{(ce)}\lambda_{S_2R}^{(c\mu)}|$                                                                   &  $\lesssim6\cdot10^{-3}$     &  \\
  $|\lambda_{S_1L,S_1R}^{(ce)}\lambda_{S_1R,S_1L}^{(c\mu)}|$                                                         &  $\lesssim6\cdot10^{-5}$     &  \\
  $|\lambda_{S_2L,S_2R}^{(ce)}\lambda_{S_2R,S_2L}^{(c\mu)}|$                                                         &  $\lesssim5\cdot10^{-5}$     &  \\
  \hline
  $|\lambda_{L,R}^{(ue)}\lambda_{L,R}^{(c\mu)}+\lambda_{L,R}^{(u\mu)}\lambda_{L,R}^{(ce)}|$                          &  $\lesssim8\cdot10^{-1}$     &  $D^0\to\mu^\pm e^\mp$  \\
  $|\lambda_{L,R}^{(ue)}\lambda_{R,L}^{(c\mu)}+\lambda_{L,R}^{(u\mu)}\lambda_{R,L}^{(ce)}|$                          &  $\lesssim1\cdot10^{-2}$     &  \\
  \hline
  $|\lambda_L^{(u\mu)}\lambda_L^{(c\mu)}+\lambda_R^{(u\mu)}\lambda_R^{(c\mu)}|$                                      &  $\lesssim6\cdot10^{-2}$     &  $D^0\to\mu^+\mu^-$  \\
  $|\lambda_L^{(u\mu)}\lambda_R^{(c\mu)}\pm\lambda_R^{(u\mu)}\lambda_L^{(c\mu)}|$                                    &  $\lesssim4\cdot10^{-3}$     &  \\
 \end{tabular}
 \caption{Scalar $SU(2)$-singlet and -doublet leptoquark constraints for real couplings scaling as $\text{TeV}/M$ and $\sqrt{\text{TeV}/M}$ for $\Delta m_{D^0}$.
 Additionally, $|\mathrm{Im}[\lambda_L^{(u\mu)}(\lambda_R^{(u\mu)})^*]|\lesssim3\cdot10^{-7}$ and $\mathrm{Im}[\lambda_L^{(ce)}(\lambda_R^{(ce)})^*]\lesssim3\cdot10^{-10}$ 
 from bounds on neutron and  electron electric dipole moments.}
 \label{tab:LQ_S12_constraints}
\end{table}
\begin{table}[!htb]
 \centering
 \begin{tabular}{c|c|c}
  couplings/mass                                                     &  constraint               &  observable  \\
  \noalign{\hrule height 1pt}
  $|\lambda^{(ue)}|$                                                 &  $\lesssim0.1$            &  $Q_w\,(\text{Cs})$  \\
  \hline
  $|\lambda^{(ue)}|$                                                 &  $\lesssim9\cdot10^{-2}$  &  $V_{ud}$  \\
  \hline
  $\lambda^{(ue)}\lambda^{(ce)}$                                     &  $\sim[-0.6,1]$           &  $D^+\to\mu^+\nu_\mu$  \\
  \hline
  $|\lambda^{(ue)}\lambda^{(ce)}|$                                   &  $\lesssim1\cdot10^{-1}$  &  $D^+\to\pi^+e^+e^-$  \\
  \hline
  $|\lambda^{(ue)}\lambda^{(c\mu)}|$                                 &  $\lesssim2\cdot10^{-1}$  &  $D^+\to\pi^+e^+\mu^-$  \\
  \hline
  $|\lambda^{(u\mu)}\lambda^{(ce)}|$                                 &  $\lesssim3\cdot10^{-1}$  &  $D^+\to\pi^+e^-\mu^+$  \\
  \hline
  $|\lambda^{(u\mu)}\lambda^{(c\mu)}|$                               &  $\lesssim1\cdot10^{-2}$  &  $D^+\to\pi^+\mu^+\mu^-$  \\
  \hline
  $|\lambda^{(u\mu)}\lambda^{(c\mu)}|$                               &  $\lesssim6\cdot10^{-2}$  &  $D^0\to\mu^+\mu^-$  \\
  \hline
  $(-|\lambda^{(ue)}|^2+|\lambda^{(u\mu)}|^2)^{1/2}$                 &  $\sim[0.2,0.4]$          &  $\Delta R_{e/\mu}$  \\
  \hline
  $|\lambda^{(ue)}\lambda^{(u\mu)}+\lambda^{(ce)}\lambda^{(c\mu)}|$  &  $\lesssim5\cdot10^{-4}$  &  $\mu^-\to e^-\gamma$  \\
  \hline
  $(\lambda^{(ue)}\lambda^{(ce)}+\lambda^{(u\mu)}\lambda^{(c\mu)})$  &  $\sim[0,0.007]$          &  $\Delta m_{D^0}$  \\
  \hline
  $|\lambda^{(ue)}\lambda^{(c\mu)}+\lambda^{(u\mu)}\lambda^{(ce)}|$  &  $\lesssim8\cdot10^{-1}$  &  $D^0\to\mu^\pm e^\mp$  \\
  \hline
  $|\lambda^{(ue)}\lambda^{(u\mu)}|$                                 &  $\lesssim7\cdot10^{-7}$  &  $\mu-e\,(\text{Au})$  \\
  $|\lambda^{(ce)}\lambda^{(c\mu)}|$                                 &  $\lesssim9\cdot10^{-3}$  &  \\
  \hline
  $|\lambda^{(ue)}\lambda^{(u\mu)}|$                                 &  $\lesssim1\cdot10^{-2}$  &  $\mu^-\to e^-e^+e^-$  \\
  $|\lambda^{(ce)}\lambda^{(c\mu)}|$                                 &  $\lesssim6\cdot10^{-3}$  &  \\
 \end{tabular}
 \caption{Scalar $SU(2)$-triplet leptoquark constraints for real couplings scaling as $\text{TeV}/M$ and $\sqrt{\text{TeV}/M}$ for $\Delta m_{D^0}$.
 For the constraint from the weak charge we apply its $2\sigma$ interval.}
 \label{tab:LQ_S3_constraints}
\end{table}
\begin{table}[!htb]
 \centering
 \begin{tabular}{c|c|c}
  couplings/mass                                                                      &  constraint               &  observable  \\
  \noalign{\hrule height 1pt}
  $|\lambda_{\tilde V_{1,2}}^{(ue)}|$                                                 &  $\lesssim0.1$            &  $Q_w\,(\text{Cs})$  \\
  $|\lambda_{V_3}^{(ue)}|$                                                            &  $\sim0.1$                &  \\
  \hline
  $|\lambda_{V_3}^{(ue)}|$                                                            &  $\lesssim6\cdot10^{-2}$  &  $V_{ud}$  \\
  \hline
  $\lambda_{V_3}^{(ue)}\lambda_{V_3}^{(ce)}$                                          &  $\sim[-0.3,0.5]$         &  $D^+\to\mu^+\nu_\mu$  \\
  \hline
  $|\lambda^{(ue)}\lambda^{(ce)}|$                                                    &  $\lesssim6\cdot10^{-2}$  &  $D^+\to\pi^+e^+e^-$  \\
  \hline
  $|\lambda^{(ue)}\lambda^{(c\mu)}|$                                                  &  $\lesssim1\cdot10^{-1}$  &  $D^+\to\pi^+e^+\mu^-$  \\
  \hline
  $|\lambda^{(u\mu)}\lambda^{(ce)}|$                                                  &  $\lesssim1\cdot10^{-1}$  &  $D^+\to\pi^+e^-\mu^+$  \\
  \hline
  $|\lambda^{(u\mu)}\lambda^{(c\mu)}|$                                                &  $\lesssim1\cdot10^{-2}$  &  $D^+\to\pi^+\mu^+\mu^-$  \\
  \hline
  $|\lambda^{(u\mu)}\lambda^{(c\mu)}|$                                                &  $\lesssim3\cdot10^{-2}$  &  $D^0\to\mu^+\mu^-$  \\
  \hline
  $(-|\lambda_{V_3}^{(ue)}|^2+|\lambda_{V_3}^{(u\mu)}|^2)^{1/2}$                      &  $\sim[0.2,0.3]$          &  $\Delta R_{e/\mu}$  \\
  \hline
  $|\lambda^{(ue)}\lambda^{(c\mu)}+\lambda^{(u\mu)}\lambda^{(ce)}|$                   &  $\lesssim4\cdot10^{-1}$  &  $D^0\to\mu^\pm e^\mp$  \\
  \hline
  $|\lambda^{(qe)}\lambda^{(q\mu)}|$                                                  &  $\lesssim1\cdot10^{-4}$  &  $\mu^-\to e^-\gamma$  \\
  $|\lambda_{V_3}^{(qe)}\lambda_{V_3}^{(q\mu)}|$                                      &  $\lesssim6\cdot10^{-5}$  &  \\
  \hline
  $|\lambda^{(ue)}\lambda^{(u\mu)}|$                                                  &  $\lesssim7\cdot10^{-7}$  &  $\mu-e\,(\text{Au})$  \\
  $|\lambda_{\tilde V_1,\tilde V_2}^{(ce)}\lambda_{\tilde V_1,\tilde V_2}^{(c\mu)}|$  &  $\lesssim1\cdot10^{-2}$  &  \\
  $|\lambda_{V_2}^{(ce)}\lambda_{V_2}^{(c\mu)}|$                                      &  $\lesssim6\cdot10^{-3}$  &  \\
  $|\lambda_{V_3}^{(ce)}\lambda_{V_3}^{(c\mu)}|$                                      &  $\lesssim7\cdot10^{-3}$  &  \\
  \hline
  $|\lambda^{(ue)}\lambda^{(u\mu)}|$                                                  &  $\lesssim4\cdot10^{-4}$  &  $\mu^-\to e^-e^+e^-$  \\
  $|\lambda_{V_3}^{(ue)}\lambda_{V_3}^{(u\mu)}|$                                      &  $\lesssim2\cdot10^{-4}$  &  \\
  $|\lambda_{\tilde V_1,\tilde V_2}^{(ce)}\lambda_{\tilde V_1,\tilde V_2}^{(c\mu)}|$  &  $\lesssim8\cdot10^{-4}$  &  \\
  $|\lambda_{V_2}^{(ce)}\lambda_{V_2}^{(c\mu)}|$                                      &  $\lesssim6\cdot10^{-4}$  &  \\
  $|\lambda_{V_3}^{(ce)}\lambda_{V_3}^{(c\mu)}|$                                      &  $\lesssim3\cdot10^{-4}$  &  \\
 \end{tabular}
 \caption{Vector leptoquark constraints for real couplings scaling as $\text{TeV}/M$.
 For the constraint on $\tilde V_{1,2}$ from $Q_w$  we apply its $2\sigma$ interval.
For $V_3$ all constraints  have to be multiplied with a factor of 1/2.}
 \label{tab:LQ_V_constraints}
\end{table}
Our constraints  are consistent with, update and extend those of \cite{Davidson:1993qk,Carpentier:2010ue} and we note that quark doublet couplings are additionally constrained by kaon physics \cite{Davidson:1993qk,Carpentier:2010ue}. Results are given in Table~\ref{tab:LQ_Kaon_constraints}.
\begin{table}[!htb]
 \centering
 \begin{tabular}{c|c|c}
  couplings/mass                                                                                      &  constraint               &  observable  \\
  \noalign{\hrule height 1pt}
  $|\lambda_{S_1L}^{(ul)}\lambda_{S_1L}^{(cl')}|$                                                     &  $\lesssim4\cdot10^{-4}$  &  $(K^+\to\pi^+\bar\nu\nu)/(K^+\to\pi^0\bar e\nu_e)$  \\
  \hline
  $|\lambda_{S_2R}^{(ue)}\lambda_{S_2R}^{(ce)}|$                                                      &  $\lesssim2\cdot10^{-3}$  &  $K_L^0\to\bar ee$  \\
  $|\lambda_{S_2R}^{(u\mu)}\lambda_{S_2R}^{(ce)}|$                                                    &  $\lesssim1\cdot10^{-5}$  &  $K_L^0\to\bar e\mu$  \\
  $|\lambda_{S_2R}^{(u\mu)}\lambda_{S_2R}^{(c\mu)}|$                                                  &  $\lesssim3\cdot10^{-4}$  &  $K_L^0\to\bar\mu\mu$  \\
  \hline
  $|\lambda_{S_3}^{(ue)}\lambda_{S_3}^{(ce)}|$                                                        &  $\lesssim4\cdot10^{-4}$  &  $(K^+\to\pi^+\bar\nu\nu)/(K^+\to\pi^0\bar e\nu_e)$  \\
  $|\lambda_{S_3}^{(ue)}\lambda_{S_3}^{(c\mu)}|\,,\quad|\lambda_{S_3}^{(u\mu)}\lambda_{S_3}^{(ce)}|$  &  $\lesssim1\cdot10^{-5}$  &  $K_L^0\to\bar e\mu$  \\
  $|\lambda_{S_3}^{(u\mu)}\lambda_{S_3}^{(c\mu)}|$                                                    &  $\lesssim3\cdot10^{-4}$  &  $K_L^0\to\bar\mu\mu$  \\
  \hline
  $|\lambda_{V_2}^{(ue)}\lambda_{V_2}^{(ce)}|$                                                        &  $\lesssim1\cdot10^{-3}$  &  $K_L^0\to\bar ee$  \\
  $|\lambda_{V_2}^{(ue)}\lambda_{V_2}^{(c\mu)}|\,,\quad|\lambda_{V_2}^{(u\mu)}\lambda_{V_2}^{(ce)}|$  &  $\lesssim5\cdot10^{-6}$  &  $K_L^0\to\bar e\mu$  \\
  $|\lambda_{V_2}^{(u\mu)}\lambda_{V_2}^{(c\mu)}|$                                                    &  $\lesssim2\cdot10^{-4}$  &  $K_L^0\to\bar\mu\mu$  \\
  \hline
  $|\lambda_{V_3}^{(ue)}\lambda_{V_3}^{(ce)}|$                                                        &  $\lesssim8\cdot10^{-5}$  &  $(K^+\to\pi^+\bar\nu\nu)/(K^+\to\pi^0\bar e\nu_e)$  \\
  $|\lambda_{V_3}^{(ue)}\lambda_{V_3}^{(c\mu)}|\,,\quad|\lambda_{V_3}^{(u\mu)}\lambda_{V_3}^{(ce)}|$  &  $\lesssim3\cdot10^{-6}$  &  $K_L^0\to\bar e\mu$  \\
  $|\lambda_{V_3}^{(u\mu)}\lambda_{V_3}^{(c\mu)}|$                                                    &  $\lesssim7\cdot10^{-5}$  &  $K_L^0\to\bar\mu\mu$  \\
 \end{tabular}
 \caption{Constraints on the leptoquark coupling products from kaon decays \cite{Carpentier:2010ue} scaling as $\text{TeV}/M$.}
 \label{tab:LQ_Kaon_constraints}
\end{table}
Next, we calculate the constraints of Tables~\ref{tab:LQ_S12_constraints}-\ref{tab:LQ_V_constraints}, where the experimental limits are given in App.~\ref{app:parameters}.
We note that fermion doublets coupling to leptoquarks are implicitly added.
We obtain constraints using  $D\to Pll$ (Eq.~(\ref{eq:dGammaq2_eob})), $D\to Pe\mu$ (Eq.~(\ref{eq:dGammaq2_DPemu})), $D\to ll$ (Eq.~(\ref{eq:B_D0ll})) and $D\to\mu e$ (Eq.~(\ref{eq:B_D0mue})).

Scalar leptoquarks contribute to the $D^0-\bar D^0$ mass difference \cite{Davidson:1993qk,Saha:2010vw}
\begin{align}
 \Delta^Sm_{D^0}=\frac23m_Df_D^2B_D\frac{\left(\lambda_{L,R}^{(cl)}\left(\lambda_{L,R}^{(ul)}\right)^*\right)^2}{64\pi^2M^2}
\end{align}
(times 2 for $S_2|_L$ and times 5 for $S_3$). While constraints from $|\Delta C|=1$ transitions scale as $\lambda \lambda^*/M^2$, the ones from mixing behave differently,
as $(\lambda \lambda^*)^2/M^2$. In our analysis we neglect the SM contribution \cite{Golowich:2005pt,Petrov:2013usa}.

Matching onto the nuclear weak charge \cite{Gresham:2012wc}
\begin{align}
 Q_w(Z,N)=-2((2Z+N)C_{1u}+(2N+Z)C_{1d})\,,
\end{align}
where $Z$ is the proton number and $N$ is the neutron number, we find
\begin{align}
 &\delta_{S_1}C_{1u}=\delta_{S_2}C_{1u}=-\frac{\sqrt 2}{8G_F}\frac{\left|\lambda_R^{(ue)}\right|^2-\left|\lambda_L^{(ue)}\right|^2}{M^2}\,,\quad\delta_{S_2}C_{1d}=-\frac{\sqrt 2}{8G_F}\frac{\left|\lambda_{S_2R}^{(ue)}\right|^2}{M^2}\,,\nonumber\\
 &\delta_{S_3}C_{1u}=\frac12\delta_{S_3}C_{1d}=\frac{\sqrt 2}{8G_F}\frac{\left|\lambda_{S_3}^{(ue)}\right|^2}{M^2}\,,\nonumber\\
 &\delta_{\tilde V_1}C_{1u}=\delta_{\tilde V_2}C_{1u}=\frac{\sqrt 2}{4G_F}\frac{\left|\lambda_{\tilde V_{1,2}}^{(ue)}\right|^2}{M^2}\,,\nonumber\\
 &\delta_{V_3}C_{1u}=-\frac{\sqrt 2}{2G_F}\frac{\left|\lambda_{V_3}^{(ue)}\right|^2}{M^2}\,,\quad\delta_{V_3}C_{1d}=-\frac{\sqrt 2}{4G_F}\frac{\left|\lambda_{V_3}^{(ue)}\right|^2}{M^2}\,.
\end{align}
We do not match $V_2$ due to an additional $d_R$-quark coupling \cite{Davidson:1993qk}.

The shift in the anomalous magnetic moment of a fermion $f$ due to a scalar LQ reads \cite{Cheung:2001ip} 
\begin{align}
 \Delta^Sa_f&=-\frac{3m_f}{16\pi^2}\frac1{M^2}\bigg(m_f\left(\left|\lambda_L^{(ff')}\right|^2+\left|\lambda_R^{(ff')}\right|^2\right)\left(\frac13Q_e^{(f')}-\frac16Q_e\right)\nonumber\\
 &+m_{f'}\mathrm{Re}\left[\lambda_L^{(ff')}\left(\lambda_R^{(ff')}\right)^*\right]\left(\left(-3-2\ln\frac{m_{f'}^2}{M^2}\right)Q_e^{(f')}-Q_e\right)\bigg)\,.
\end{align}
Here, $Q_e$ and $Q_e^{(f')}$ denote the electric charges of the leptoquark and the fermion $f^\prime$  in the loop, respectively.
The contribution to  the electric dipole moment  of $f$ reads  \cite{Cheung:2001ip}
\begin{align} 
 d_f=\frac e{32\pi^2}\frac1{M^2}m_{f'}\mathrm{Im}\left[\lambda_L^{(ff')}\left(\lambda_R^{(ff')}\right)^*\right]\left(\left(-3-2\ln\frac{m_{f'}^2}{M^2}\right)Q_e^{(f')}-Q_e\right) \, ,
\end{align}
times 3 for color if $f'$ is a quark. 
For electrons $|d_e^{\rm SM}| \leq 10^{-38} \,e\;\text{cm}$  \cite{Pospelov:1991zt}. The neutron electric dipole moment receives contributions from quarks
$d_n=4/3d_d-1/3d_u$, with $d_n^\text{SM}\sim\mathcal O(10^{-34})\,e\;\text{cm}$ \cite{Czarnecki:1997bu}.

The lepton flavor violating radiative muon decay in case of a scalar LQ  is \cite{Benbrik:2008si}
\begin{align}
 \delta_S\mathcal B_{\mu\to e\gamma}=\frac{\alpha_e}{4\Gamma_\mu}m_\mu^5\left(\left|F_{2LR}^\gamma\right|^2+\left|F_{2RL}^\gamma\right|^2\right)\,,
\end{align}
where we note the typo $F_1\leftrightarrow F_2$ in \cite{Benbrik:2008si}
\begin{align}\label{eq:F2gamma}
 F_{2LR,2RL}^\gamma&=\frac{3}{16\pi^2}\frac1{M^2}\bigg(\lambda_{L,R}^{(q\mu)}\left(\lambda_{L,R}^{(qe)}\right)^*\left(\frac16Q_e^{(q)}-\frac1{12}Q_e\right)\nonumber\\
 &-\frac{m_q}{m_\mu}\lambda_{R,L}^{(q\mu)}\left(\lambda_{L,R}^{(qe)}\right)^*\left(\left(-\frac32-\ln\frac{m_q^2}{M^2}\right)Q_e^{(q)}-\frac12Q_e\right)\bigg)\,.
\end{align}
In case of a vector LQ \cite{Davidson:1993qk}
\begin{align}
 \delta_V\mathcal B_{\mu\to e\gamma}=\frac1{\Gamma_\mu}\frac{m_\mu}{8\pi}\left(\left|F^V\right|^2+\left|F^A\right|^2\right)
\end{align}
with
\begin{align}\label{eq:FVA}
 \left|F^{V,A}\right|=\sqrt{\alpha_e4\pi}\frac{\left|\lambda^{(qe)}\lambda^{(q\mu)}\right|}{32\pi^2}\frac{m_\mu^2}{M^2}\left(Q_e^{(q)}2+Q_e\frac52\right)
\end{align}
(times 2 for up-type quarks in scenario $V_3$).

The lepton flavor violating muon decay in case of a scalar LQ is \cite{Benbrik:2010cf}
\begin{align}
 \delta_S\mathcal B_{\mu\to eee}&=\frac{\alpha_e^2m_\mu^5}{32\pi\Gamma_\mu}\bigg(|T_{1L}|^2+|T_{1R}|^2+\frac23\left(|T_{2L}|^2+|T_{2R}|^2\right)\left(8\ln\frac{m_\mu}{2m_e}-11\right)\nonumber\\
 &-4\mathrm{Re}[T_{1L}T_{2R}^*+T_{2L}T_{1R}^*]\nonumber\\
 &+\frac13\left(2\left(|Z_L g_{Ll}|^2+|Z_R g_{Rl}|^2\right)+|Z_L g_{Rl}|^2+|Z_R g_{Ll}|^2\right)\nonumber\\
 &+\frac16\left(|B_{1L}|^2+|B_{1R}|^2\right)+\frac13\left(|B_{2L}|^2+|B_{2R}|^2\right)\nonumber\\
 &+\frac23\mathrm{Re}[T_{1L}B_{1L}^*+T_{1L}B_{2L}^*+T_{1R}B_{1R}^*+T_{1R}B_{2R}^*]\nonumber\\
 &-\frac43\mathrm{Re}[T_{2R}B_{1L}^*+T_{2L} B_{1R}^*+T_{2L}B_{2R}^*+T_{2R} B_{2L}^*]\nonumber\\
 &+\frac23\mathrm{Re}[B_{1L}Z_L^*g_{Ll}+B_{1R}Z_R^*g_{Rl}+B_{2L}Z_L^*g_{Rl}+B_{2R}Z_R^*g_{Ll}]\nonumber\\
 &+\frac23\mathrm{Re}[2(T_{1L}Z_L^*g_{Ll}+T_{1R}Z_R^*g_{Rl})+T_{1L}Z_L^*g_{Rl}+T_{1R}Z_R^*g_{Ll}]\nonumber\\
 &+\frac23\mathrm{Re}[-4(T_{2R}Z_L^*g_{Ll}+T_{2L}Z_R^*g_{Rl})-2(T_{2L}Z_R^*g_{Ll}+T_{2R}Z_L^*g_{Rl})]\bigg)\,,
\end{align}
where we correct the typo in $Z_{L,R}$ related terms in \cite{Benbrik:2010cf}
\begin{align}
 T_{1L,1R}&=-\frac3{16\pi^2}\frac1{M^2}\lambda_{L,R}^{(q\mu)}\left(\lambda_{L,R}^{(qe)}\right)^*\left(\left(\frac49+\frac13\ln\frac{m_q^2}{M^2}\right)Q_e^{(q)}+\frac{-1}{18}Q_e\right)\,,\\
 T_{2L,2R}&=\frac3{16\pi^2}\frac1{M^2}\bigg(-\left(\lambda_{R,L}^{(q\mu)}\left(\lambda_{R,L}^{(qe)}\right)^*\frac16+\frac{m_q}{m_\mu}\lambda_{R,L}^{(q\mu)}\left(\lambda_{L,R}^{(qe)}\right)^*\left(-\frac32-\ln\frac{m_q^2}{M^2}\right)\right)Q_e^{(q)}\nonumber\\
 &+\left(\lambda_{R,L}^{(q\mu)}\left(\lambda_{R,L}^{(qe)}\right)^*\frac1{12}+\frac{m_q}{m_\mu}\lambda_{R,L}^{(q\mu)}\left(\lambda_{L,R}^{(qe)}\right)^*\frac12\right)Q_e\bigg)\,,\\
 Z_{L,R}&=-\frac3{16\pi^2}\frac1{M^2}\lambda_{L,R}^{(q\mu)}\left(\lambda_{L,R}^{(qe)}\right)^*\frac1{m_Z^2\sin^2\theta_w\cos^2\theta_w}\nonumber\\
 &\times\left(m_\mu^2\frac382g_{Lq,Rq}-m_q^2\left(1+\ln\frac{m_q^2}{M^2}\right)g_{Rq,Lq}+m_\mu^2\frac382(-g)\right)\,,\\
 B_{1L,1R}&=\frac3{32\pi^2}\lambda_{L,R}^{(q\mu)}\left(\lambda_{L,R}^{(qe)}\right)^*\left|\lambda_{L,R}^{(q'e)}\right|^2\frac{-1}{M^2}\,,\\
 B_{2L,2R}&=\frac3{64\pi^2}\lambda_{L,R}^{(q\mu)}\left(\lambda_{L,R}^{(qe)}\right)^*\left|\lambda_{R,L}^{(q'e)}\right|^2\frac{-1}{M^2}
\end{align}
and
\begin{align}
 g_{Lf,Rf}=T_3^{(f_L,f_R)}-Q_e^{(f)}\sin^2\theta_w\,,\quad g=T_3-Q_e\sin^2\theta_w
\end{align}
($g_L\leftrightarrow g_R$ for $S_2$).
Here, $T_3$ is the third component of the weak isospin of the LQ and $f_{L,R}$ label chiral fermions, that is, $g_{Lf,Rf}$ are SM couplings and $g$ is the LQ coupling.
In case of a vector LQ \cite{Gvozdev:1994bk}
\begin{align}
 \delta_V\mathcal B_{\mu\to eee}=\frac{3\alpha_e^2}{8\pi^2}\left(Q_e^{(q)}\right)^2\ln^2\frac{m_q^2}{M^2}\frac{\left|\lambda^{(q\mu)}\lambda^{(qe)}\right|^2}{G_F^2M^4}
\end{align}
(times 4 for up-type quarks in scenario $V_3$), where we neglect terms $\sim Q_e$, $m_f^2/M^2$-suppressed box-diagrams and $m_f^2/m_Z^2$-suppressed $Z$-diagrams.

Matching onto the $\mu-e$ conversion in nuclei rate \cite{Kitano:2002mt}
\begin{align}
 \Gamma_{\mu-e}&=4m_\mu^5\left|\frac14C_{DR}D+C_{SL}G_S^{(u,p)}S^{(p)}+C_{SL}G_S^{(u,n)}S^{(n)}+2C_{VR}V^{(p)}+C_{VR}V^{(n)}\right|^2\nonumber\\
 &+4m_\mu^5\left|\frac14C_{DL}D+C_{SR}G_S^{(u,p)}S^{(p)}+C_{SR}G_S^{(u,n)}S^{(n)}+2C_{VL}V^{(p)}+C_{VL}V^{(n)}\right|^2
\end{align}
we find
\begin{align}
 &\delta_SC_{VR}=-\frac{\lambda_R^{(ue)}\left(\lambda_R^{(u\mu)}\right)^*}{4M^2}\,,\quad\delta_SC_{VL}=-\frac{\lambda_L^{(ue)}\left(\lambda_L^{(u\mu)}\right)^*}{4M^2}\,,\nonumber\\
 &\delta_SC_{SR}=-\frac{\lambda_L^{(ue)}\left(\lambda_R^{(u\mu)}\right)^*}{4M^2}\,,\quad\delta_SC_{SL}=-\frac{\lambda_R^{(ue)}\left(\lambda_L^{(u\mu)}\right)^*}{4M^2}\,,\nonumber\\
 &\delta_SC_{DR,DL}=\frac1{2m_\mu}F_{2RL,2LR}^\gamma\,,\nonumber\\
 &\delta_{\tilde V_1}C_{VR}=\frac{\lambda_{\tilde V_1}^{(ue)}\left(\lambda_{\tilde V_1}^{(u\mu)}\right)^*}{2M^2}\,,\quad\delta_{V_2}C_{VR}=\frac{\lambda_{V_2}^{(u\mu)}\left(\lambda_{V_2}^{(ue)}\right)^*}{2M^2}\,,\nonumber\\
 &\delta_{\tilde V_2}C_{VL}=\frac{\lambda_{\tilde V_2}^{(u\mu)}\left(\lambda_{\tilde V_2}^{(ue)}\right)^*}{2M^2}\,,\quad\delta_{V_3}C_{VL}=\frac{\lambda_{V_3}^{(ue)}\left(\lambda_{V_3}^{(u\mu)}\right)^*}{M^2}\,,\nonumber\\
 &\delta_{\tilde V_1,V_2}C_{DR,DL}=-\frac1{4m_\mu^2\sqrt{\alpha_e4\pi}}\left(\left|F^V\right|\pm\left|F^A\right|\right)\,,\nonumber\\
 &\delta_{\tilde V_2,V_3}C_{DR,DL}=-\frac1{4m_\mu^2\sqrt{\alpha_e4\pi}}\left(\left|F^V\right|\mp\left|F^A\right|\right)\,,
\end{align}
where $F_{2RL,2LR}^\gamma$ are given by Eq.~(\ref{eq:F2gamma}) and $|F^{V,A}|$ are given by Eq.~(\ref{eq:FVA}) (times 2 for up-type quarks in scenario $V_3$).
We neglect loop-suppressed gluonic interactions.
The nucleon form factors are given as $G_S^{(u,p)}=5.1$ and $G_S^{(u,n)}=4.3$ \cite{Kitano:2002mt} and we take the overlap integrals of muons and electrons weighted by proton and neutron densities for titanium and gold
\begin{align}
 &D_\text{Ti}=0.0864\,,\quad S_\text{Ti}^{(p)}=0.0368\,,\quad S_\text{Ti}^{(n)}=0.0435\,,\quad V_\text{Ti}^{(p)}=0.0396\,,\quad V_\text{Ti}^{(n)}=0.0468\,,\nonumber\\
 &D_\text{Au}=0.189\,,\quad S_\text{Au}^{(p)}=0.0614\,,\quad S_\text{Au}^{(n)}=0.0918\,,\quad V_\text{Au}^{(p)}=0.0974\,,\quad V_\text{Au}^{(n)}=0.146\,.
\end{align}

Matching onto the leptonic pseuodoscalar decay rate \cite{Barranco:2014bva}
\begin{align}
 \Gamma_{P\to l\nu}=\frac{G_F^2f_P^2\left(m_P^2-m_l^2\right)^2}{8\pi m_P^3}\left|m_lV_{qq'}+m_l\frac{\sqrt2}{4G_F}(C_{VRL}-C_{VLL})+\frac{m_P^2}{m_q+m_{q'}}\frac{\sqrt2}{4G_F}(C_{SLR}-C_{SRR})\right|^2
\end{align}
we find
\begin{align}
 &\delta_{S_1}C_{VLL}=\frac12\frac{\lambda_{S_1L}^{(ql)}\left(\lambda_{S_1L}^{(q'l)}\right)^*}{M^2}\,,\quad\delta_{S_1}C_{SRR}=-\frac12\frac{\lambda_{S_1R}^{(ql)}\left(\lambda_{S_1L}^{(q'l)}\right)^*}{M^2}\,,\nonumber\\
 &\delta_{S_2}C_{SRR}=\frac12\frac{\lambda_{S_2R}^{(q'l)}\left(\lambda_{S_2L}^{(ql)}\right)^*}{M^2}\,,\nonumber\\
 &\delta_{S_3}C_{VLL}=-\frac12\frac{\lambda_{S_3}^{(ql)}\left(\lambda_{S_3}^{(q'l)}\right)^*}{M^2}\,,\nonumber\\
 &\delta_{V_3}C_{VLL}=-\frac{\lambda_{V_3}^{(q'l)}\left(\lambda_{V_3}^{(ql)}\right)^*}{M^2}\,,
\end{align}
We do not match $V_2$ due to an additional $d_L$-quark coupling \cite{Davidson:1993qk}.

We deduce the shift in $R_{e/\mu}=\Gamma_{\pi\to e\nu_e}/\Gamma_{\pi\to\mu\nu_\mu}$
\begin{align}
 \delta_{LQ}R_{e/\mu}&=R_{e/\mu}^\text{SM}\frac1{\sqrt2G_FV_{ud}}\mathrm{Re}\bigg[\left(C_{VRL}^{(l=e)}-C_{VLL}^{(l=e)}\right)-\left(C_{VRL}^{(l=\mu)}-C_{VLL}^{(l=\mu)}\right)\nonumber\\
 &+\frac{m_\pi^2}{m_u+m_d}\left(\frac{C_{SLR}^{(l=e)}-C_{SRR}^{(l=e)}}{m_e}-\frac{C_{SLR}^{(l=\mu)}-C_{SRR}^{(l=\mu)}}{m_\mu}\right)\bigg]\,.
\end{align}
and the shift in the CKM parameter
\begin{align}
 \frac{\Delta^{LQ}V_{ud}}{V_{ud}}=\frac{\sqrt2}{4G_F}C_{VLL}^{(ue)}
\end{align}
by means of quark beta decay normalized to muon decay.

We match onto nuclear beta decay parameters to constrain Wilson coefficients \cite{Wauters:2013loa}
\begin{align}
 -0.14\cdot10^{-2}<\frac{G_F\alpha_e}{\sqrt2\pi}\frac{C_T+C_{T5}}{C_A}<1.4\cdot10^{-2}\quad(90\%\,\text{C.L.})\,,
\end{align}
where $C_A^\text{SM}=-1.27\,G_FV_{ud}$.

We obtain no constraints better than  $|\lambda| \lesssim M/{\rm TeV}$ from the decay $\pi^0\to\mu e$, $\Delta m_D$ via vector LQs \cite{Inami:1980fz}, the $D^0-\bar D^0$ lifetime difference \cite{Hagelin:1981zk,Dighe:2007gt}, the anomalous magnetic moment via vector LQs \cite{Leveille:1977rc}, the decay $Z\to ff$ via scalar LQs \cite{Bhattacharyya:1994ig}, the decay $Z\to e\mu$ via  scalar LQs \cite{Benbrik:2008si}, triple correlation coefficients in nuclear beta decay \cite{Chupp:2012ta,Kozela:2009ni,Huber:2003gr} nor additional nuclear beta decay parameters \cite{Adelberger:1999ud}.

\end{document}